\documentclass[preprint,twocolumn]{aastex631}
\usepackage{graphicx,amsmath}

\usepackage{csquotes}
\usepackage{subfigure}

\newcommand{\beq}{\begin{equation}}
\newcommand{\eeq}{\end{equation}}
\newcommand{\bga}{\begin{gathered}}
\newcommand{\ega}{\end{gathered}}

\newcommand\ba{\begin{eqnarray}}
\newcommand\ea{\end{eqnarray}}
\newcommand\bea{\begin{eqnarray}}
\newcommand\eea{\end{eqnarray}}

\newcommand\be{\begin{equation}}
\newcommand\ee{\end{equation}}

\newcommand{\rmn}{\mathrm}

\def\={\nonumber &=}

\def\({\left(}
\def\){\right)}
\def\[{\left[}
\def\]{\right]}
\def\<{\left\langle}
\def\>{\right\rangle}

\def\curl{\mathcal}

\def\eq{\begin{eqnarray}}
\def\qe{\end{eqnarray}}

\def\and{\quad \mbox{and} \quad}

\def\bfnl{\kern2pt\overline{\kern-2ptf}_\textrm{NL}}

\def\exp{\textrm{exp}}

\def\Q{\curl{Q}}

\def\R{\curl{R}}

\def\barQ{\kern2pt\overline{\kern-2pt\curl{Q}}}

\def\bargamma{\kern2pt\overline{\kern-2pt\gamma}}
\def\barzeta{\kern2pt\overline{\kern-2pt\zeta}}

\def\barR{\kern2pt\overline{\kern-2pt\curl{R}}}

\def\eqref#1{(\ref{#1})}

\def\leaderfi1{\leaders\hbox to 5pt{\hss.\hss}\hfil}

\def\setsize{\csname @setfontsize\endcsname \setsize}

\newcommand{\uvec}[1]{\hat{\boldsymbol{#1}}}

\begin{document}

\title{Probing the circumgalactic medium with CMB polarization statistical anisotropy}
\author{Anirban Roy} 
\affiliation{Department of Astronomy, Cornell University, Ithaca, New York 14853, USA}

\author{Alexander van Engelen}
\affiliation{School of Earth and Space Exploration, Arizona State University, Tempe, AZ 85287, USA}

\author{Vera Gluscevic}
\affiliation{Department of Physics \& Astronomy, University of Southern California, Los Angeles, CA, 90007, USA}

\author{Nicholas Battaglia}
\affiliation{Department of Astronomy, Cornell University, Ithaca, New York 14853, USA}

\begin{abstract}
As cosmic microwave background (CMB) photons traverse the Universe, anisotropies can be induced via Thomson scattering (proportional to the integrated electron density; optical depth) and inverse Compton scattering (proportional to the integrated electron pressure; thermal Sunyaev-Zel'dovich effect). Measurements of anisotropy in optical depth $\tau$ and Compton $y$ parameter are imprinted by the galaxies and galaxy clusters and are thus sensitive to the thermodynamic properties of circumgalactic medium and intergalactic medium.
We use an analytic halo model to predict the power spectrum of the optical depth ($\tau\tau$), the cross-correlation between the optical depth and the Compton $y$ parameter ($\tau y$), as well as the cross-correlation between the optical depth and galaxy clustering ($\tau g$), and compare this model to cosmological simulations. We constrain the optical depths of halos at $z\lesssim 3$ using a technique originally devised to constrain patchy reionization at a much higher redshift range. 
The forecasted signal-to-noise ratio is 2.6, 8.5, and 13, respectively, for a CMB-S4-like experiment 
and a VRO-like optical survey. We show that a joint analysis of these probes can constrain the amplitude of the density profiles of halos to 6.5\% and the pressure profile to 13\%, marginalizing over the outer slope of the pressure profile. These constraints translate to astrophysical parameters related to the physics of galaxy evolution, such as the gas mass fraction, $f_{\rm g}$, which can be constrained to 5.3\% uncertainty at $z\sim 0$, assuming an underlying model for the shape of the density profile. The cross-correlations presented here are complementary to other CMB and galaxy cross-correlations since they do not require spectroscopic galaxy redshifts and are another example of how such correlations are a powerful probe of the astrophysics of galaxy evolution.

\end{abstract}

\section {Introduction}\label{sec:introduction}
Current-generation cosmic microwave background (CMB) observations by the \textit{Planck} satellite and various Stage-3 ground-based experiments have mapped the temperature anisotropy with unprecedented precision, saturating the cosmic variance limit on a wide range of angular scales \citep{P18:main, ACT:2020gnv, Reichardt:2020}.
They have, however, only begun to tap the information encoded within the polarization and lensing anisotropy, leaving much to be accomplished by future experiments.
The next-generation ground-based observatories such as Simons Observatory\footnote{\href{https://simonsobservatory.org/}{https://simonsobservatory.org/}} \citep{SimonsObservatory}, CMB S4\footnote{\href{https://cmb-s4.org/}{https://cmb-s4.org/}} \citep{CMBS4}, and SPT-3G\footnote{\href{https://pole.uchicago.edu/}{https://pole.uchicago.edu/}} \citep{SPT-3G:2021vps} are envisioned to realize a dramatic leap forward in terms of polarization and lensing anisotropy measurements on all angular scales, optimizing the science goals tied to the effects on the CMB-power-spectra damping tail: neutrino mass measurements, searches for light relic particles, dark matter, etc.
The discovery potential of high-resolution polarization measurements, however, goes beyond the damping-tail science.
In particular, many physical effects produce non-Gaussian footprints in CMB maps, which can be sought with higher-order statistical-anisotropy estimators.
In this work, we focus on non-Gaussian signatures of secondary anisotropy imprinted by the inhomogeneous distribution of free electrons in halos at $z\lesssim 3$.

The gas present within the virial radii of galaxies and outside their disks is known as the circumgalactic medium (CGM). The CGM connects the interstellar medium (ISM) to the intergalactic medium (IGM), and is affected by the physics of the radiative process of galactic winds, active galactic nuclei (AGN) and supernovae feedback \citep{Werk:2014fza, 2015ApJ...812...83N, 2017ARA&A..55..389T}. These processes change the pressure and density profile of galaxies, and hence the detection of thermal Sunyaev Zel'dovich (tSZ) from CGM is a potential probe to help understanding the thermodynamics of halos. Thus, CMB observations can provide a new window into studying the CGM. Several cross-correlation studies such as tSZ-galaxy clustering and tSZ-CMB lensing are also useful to measure the pressure of gas and their evolution in host halos \citep{Hill:2013dxa, Battaglia:2014era, 2015A&A...576A..90H,  Hojjati:2016nbx,   Pandey:2019, Koukoufilippas:2019ilu,  Amodeo:2020mmu, Yan:2021gfo, Pandey:2021bdj}. The tSZ-galaxy clustering cross-correlation has been extensively studied in various aspects to measure the gas mass fraction, the gas temperature, and their evolution \citep{2015ApJ...808..151G, Singh:2015, Amodeo:2020mmu, Moser2021, 2021ApJ...913...88M}. Furthermore, the measurements of the kinetic Sunyaev Zel'dovich (kSZ) effect can provide constraints on the density profiles of gas in halos, and joint analysis of kSZ and tSZ measurements can break the degeneracies between the parameters. In addition, cross-correlations with spectroscopic galaxy samples can help us to measure the temperature profile of the halos \citep{Amodeo:2020mmu}. 

The spatial distribution of free electrons in the Universe is inhomogeneous, and the largest contribution to the statistical variance of these fluctuations is generated during the epoch of reionization (EoR) since the reionization process is highly non-Gaussian in nature. The small scale temperature and polarization anisotropies in CMB are suppressed by a factor of $e^{-\tau(\uvec{n})}$ due to effects of \enquote{patchy screening}, where $\tau(\uvec{n})$ is the direction-dependent optical depth \citep{Dvorkin:2008:tau-est, Gluscevic:2012:tau}. This can be probed by the reconstruction of optical depth field \citep{Dvorkin:2008:tau-est, Gluscevic:2012:tau, Namikawa:2017:plktau, Roy:2018, Roy:2020cqn, Guzman:2021nfk}. Similar to but smaller than the reionization process, galaxies and galaxy clusters at low redshift contribute to these optical depth fluctuations and provide a probe of the distribution of free electrons inside such halos.
Although the imprints of halos and the reionization process in CMB are related to different physical phenomena, their common trait is a qualitatively similar statistical-anisotropy signature in CMB maps that requires the same quadratic-estimator techniques to uncover. 
Even though the estimators for the reconstruction of optical depth fluctuations \citep{Dvorkin:2008:tau-est} was first developed to study the reionization process, we adopt it here to
perform a statistical study of the CGM in the low redshift Universe.

Quadratic estimators for reconstructing lensing potential have been widely used to map the matter distribution in the Universe \citep{P18:phi}, and an attempt to probe the electron density fluctuations was made by implementing quadratic optical depth estimator to the \textit{Planck} data \citep{Namikawa:2017:plktau}. As secondary anisotropies due to the fluctuations in electron density are subdominant to the lensing signal, neglecting the higher-order corrections in lensing and optical depth can bias the reconstruction noise of an estimator. A bispectrum approach was adopted to extract information about the CMB lensing and optical depth simultaneously \citep{Feng:2018eal}. Their forecast shows the cross-correlation between the reconstructed optical depth and lensing can be detected with a few $\sigma$ by upcoming CMB experiments. Furthermore, the forecasted detectability of the cross-correlation between optical depth reconstructed for a CMB-S4-like experiment and galaxy number-count for a Vera C. Rubin Observatory (VRO)-like experiment is at the order of $\sim 8\sigma$ \citep{Feng2018}. 

The measurements of density profiles are particularly important to break the degeneracy between optical depth and peculiar radial velocity in kSZ measurements \citep[e.g.,][]{Alonso2016,Madhavacheril:2019buy}. 
Other kSZ estimators such as the projected fields estimator \citep{Hill2016,Simo2016,Kusiak2021}, the tomographic kSZ estimator \citep{Shao2016}, and the pairwise kSZ estimator \citep[e.g.,][]{DeB2017,Calafut2021,Vava2021} can further provide integral constraints optical depth of halos (the baryon abundance).
Though the statistical properties of the optical depth of halos can be inferred only from such kSZ measurements, these cross-correlation techniques reduce various systematic effects.

In this paper, we propose a set of independent probes such as cross-correlation between the optical depth and galaxy clustering, as well as the cross-correlation between the optical depth and the Compton $y$ parameter to understand the gas physics in CGM. We thus present sensitivity forecasts for a CMB-S4-like experiment for each of these physical effects and explore survey strategies that maximize the quality of their measurement. We show how future measurements of these signals can constrain the astrophysical parameters of the CGM. Thus, we illustrate the broad range of CGM information available from careful statistical anisotropy measurements with next-generation CMB observations.

This paper is organized as follows. In Section \ref{sec:halomodel}, we describe the halo model approach for the calculation of several auto power spectra of optical depth, galaxy overdensity, Compton y, and cross power spectrum between them. We compare our model to simulations in Section \ref{sec:simulation}. In Section \ref{sec:noise_model}, we describe the noise model for optical depth estimator, tSZ measurements, and galaxy clustering. We present the key results of our paper in Section \ref{sec:result}, and then we present the constraints on astrophysical parameters in Section \ref{sec:fisher}. We finally draw our conclusions in Section \ref{sec:discussion}. Throughout this work, we assume the best-fit $\Lambda$CDM cosmological parameters derived from Planck TT, TE, EE+lowE+lensing signals \citep{P18:main}. 

\section{Halo Model} \label{sec:halomodel}
In this section, we review and summarize the basic formalism of the halo-model approach to calculate the auto and cross power spectra of different observables such as tSZ, optical depth, and galaxy fields. 

\subsection{Optical depth}
The integrated electron density along the line of sight is referred to as the optical depth. As the underlying density field is perturbed, it introduces spatial fluctuations in the electron distribution. This leads to the direction dependence of the optical depth, which can be written as:
\be
\tau(\uvec{n})=\sigma_{\rm T}\int a\,n_{\rm e}(\uvec{n},\chi) d\chi \,.
\label{eq:tau}
\ee 
Here, $n_{\rm e}(\uvec{n},\chi)$ is the electron density along the direction $\uvec{n}$ at a comoving distance $\chi$, $\sigma_{\rm T}$ is the Thomson scattering cross section, and $a$ is the scale factor of the Universe. The CMB temperature and polarization anisotropy are damped as
\be
\Delta T(\uvec{n})= \Delta \tilde{T}(\uvec{n})e^{-\tau(\uvec{n})}
\label{eq:primary_temp}
\ee 

\be
\Delta (Q\pm iU)(\uvec{n})= \Delta (\tilde{Q}\pm i\tilde{U})(\uvec{n})e^{-\tau(\uvec{n}})\,,
\label{eq:q_and_u}
\ee 
where $\tilde{T}$ is the temperature, $\tilde{Q}$ and $\tilde{U}$ are the Stokes parameters of CMB polarization at the last scattering surface. 

The perturbation on the free electron density $\Delta n_{\rm e} (\uvec{n}, \chi)$ introduces the perturbation $\Delta \tau(\uvec{n})$ on the sky averaged optical depth $\bar{\tau}$ as

\be
\tau(\uvec{n})=\bar{\tau}+ \Delta \tau(\uvec{n})\,.
\ee
We now proceed to relate the optical depth to the distribution of electrons within halos. To model the electron density within halos, we adopt the fitting formula derived from cosmological simulations for the gas density from \citet{Battaglia:2016xbi} as
\be
\rho_{\rm fit}(x)= \rho_0 (x/x_c)^{\gamma}\left[ 1+(x/x_c)^{\alpha}\right]^{-\left(\frac{\beta_{\rho}+\gamma}{\alpha}\right)}\,, \label{eq:rhofit}
\ee
where $x=r/r_s$, $\rho_0$ is the amplitude of density profile, $\alpha$ is the intermediate slope, and $\beta_{\rho}$ is the power law index. Then, the average stacked profile of the gas density becomes $\rho_{3D}=f_{\rm b}\rho_{\rm fit}\rho_{\rm crit}(z)$, where $f_b$ is the baryon fraction of the Universe, $\Omega_{\rm b}/\Omega_{\rm M}$, and $\rho_{\rm crit}$ is the critical density of the Universe. Thoughout this work, we keep fixed $\gamma=-0.2$ \citep{Battaglia2012}. The dependence of the parameters $\rho_0$, $\alpha$, and $\beta_{\rho}$ on the redshift and mass of the halo is fitted with a generic form of equation as:
\be
X= A_X \left( \frac{M_{200}}{10^{14}\, M_\odot}\right)^{\alpha_{m,X}}(1+z)^{\alpha_{z,X}}\,.\label{eq:params_scale}
\ee
In the above equation, index $X$ refers to the parameters of our choice related to density profiles, $\rho_0$, $\alpha$, and $\beta_{\rho}$. $A_X$ is the amplitude of the parameter, $\alpha_{m,X}$, and $\alpha_{z,X}$ are the power-law indices for the mass and redshift scaling, respectively. The values of such  fitted parameters are given in the Table \ref{table:density_params}

\begin{table}[h]
\centering
\begin{tabular}{cccc}
\hline\hline
Parameter & $A$ & $\alpha_m$ & $\alpha_z$ \\
\hline
$\rho_0$ & $4\times 10^3$ & 0.29 & -0.66 \\
$\alpha$ & 0.88 & -0.03 & 0.19 \\
$\beta_\rho$ & 3.83 & 0.04 & -0.025 \\
\hline
\end{tabular}
\caption{The parameters fitted for the density profile of halos with AGN feedback \citep{Battaglia:2016xbi}.}
\label{table:density_params}
\end{table}

Now, the Fourier transform of $\tau$ under the Limber approximation becomes \citep{Komatsu1999, Hill:2013},
\be
K^{\tau}_\ell(M,z)=\frac{4\pi r_s \sigma_T}{\ell^2_s} \int dx \frac{\sin(\ell x/\ell_s)}{\ell x/\ell_s} \rho_{\rm 3D}(x, M, z) x^2 \,,\label{eq:tau_ell}
\ee
where $K^{\tau}_\ell(M,z)$ is the kernel of the anisotropies of $\tau$ related to the density profiles of a halo and $\ell_s$ is the multipole moment corresponding to the characteristic scale radius $r_s$, of the $\rho_{\rm 3D}$ profile. The effective bias of a field is related to the bias of a dark matter halo. The effective bias of the optical depth profile of such a halo can be written as:
\be
b_{\ell}^{\rm \tau} (M, z)=  \int_{M_{\rm min}}^{M_{\rm max}} dM\frac{dn (M,z)}{dM} K^{\tau}_\ell(M,z) b(M,z) \,,
\label{eq:b_ell_tau}
\ee
where $b(M,z)$ is the halo bias and $dn(M,z)/dM$ is the halo mass function. We use the Tinker halo mass function throughout this work \citep{Tinker:2008}.
\subsection{Compton y parameter}
The CMB photons get scattered off the hot electron gas present in the halos, which causes additional anisotropies in CMB temperature \citep{Sunyaev1970, Sunyaev1972}. The temperature fluctuation due to the tSZ effect can be expressed as 
\be
\frac{\Delta T_{\rm tSZ}(\nu, \uvec{n})}{T_{\rm CMB}}=g(\nu)y(\uvec{n})\,.
\ee
In the above equation $T_{\rm CMB}$ is the sky-averaged temperature and $\Delta T_{\rm tSZ}(\nu)$ is change of temperature with respect to $T_{\rm CMB}$ at frequency $\nu$. The term $g(\nu)$, describes how the temperature fluctuation due to tSZ effect changes with frequency. In the non-relativistic limit, $g(\nu)=x\coth(x/2)-4$, where $x={h\nu}/k_{\rm B} T_{\rm e}$, where $T_{\rm e}$ is the temperature of the electron gas, $h$ is the Planck constant and $K_{\rm B}$ is the Boltzman constant. We neglect the relativistic corrections to the spectral shape of the tSZ signal \citep{Sz_relativistic, 2012MNRAS.426..510C}.  

The Compton $y$ parameter in a direction $\uvec{n}$ is given by
\be
    y(\uvec{n}) = \frac{\sigma_{\rm T}}{m_{\rm e}c^2}
    \int P_{\rm e}(\uvec{n},\chi) d\chi \,.
    \label{eq:yparam}
\ee 
Here, $c$ is the speed of light in free space and $m_{\rm e}$ is the mass of an electron. $P_{\rm e}(\uvec{n}, \chi)$ is the pressure along the direction $\uvec{n}$ at a distance $\chi$. For an isothermal medium, the Compton $y$ parameter is proportional to the optical depth of that medium. 

To calculate the $y$ parameter, we use the analytic pressure profiles described in \citet{Battaglia2012} that use a GNFW profile \citep{Zhao1996} to model the pressure. If $r$ is the radius of a halo and $r_s$ is the characteristic scale of such a halo, then the pressure fitted to the generalized NFW profile is 
\be
{P}_{\rm e}(x)= P_0 (x/x_c)^{\gamma}\left[ 1+(x/x_c)^{\alpha}\right]^{-\beta_{\rm P}}\,. 
\label{eq:pfit}
\ee
Here, $P_0$ is the amplitude of pressure, and $x_c$ is the fractional core scale radius, and $\beta_{P}$ is the power-law index. $\alpha$ and $\gamma$ are two free parameters of the model. In equation Eq.\,\ref{eq:params_scale}, index $X$ refers to the parameters of our choice related to pressure profiles, $P_0$, $x_c$, and $\beta_{P}$. The best-fit values of these parameters are given in the Table\,\ref{table:pressure_params}.

\begin{table}[h]
\centering
\begin{tabular}{cccc}
\hline\hline
Parameter & $A$ & $\alpha_m$ & $\alpha_z$ \\
\hline
$P_0$ & 18.1 & 0.154 & -0.758 \\
$x_c$ & 0.497 & -0.00865 & 0.731 \\
$\beta_P$ & 4.35 & 0.0393 & 0.415 \\
\hline
\end{tabular}
\caption{The parameters fitted for the pressure profile of halos \citep{Battaglia2012}.}
\label{table:pressure_params}
\end{table}

Now, the Fourier transform of the $y$ parameter under the Limber approximation becomes \citep{Komatsu1999, Hill:2013},:
\be
K^y_\ell(M,z)=\frac{4\pi r_s}{\ell^2_s} \int dx \frac{\sin(\ell x/\ell_s)}{\ell x/\ell_s} y(x, M, z) x^2\,. \label{eq:y_ell}
\ee

The effective bias of the $y$ parameter is given by \citep{Pandey:2019}:
\be
b_{\ell}^{y} (M, z)=  \int_{M_{\rm min}}^{M_{\rm max}} dM\frac{dn (M,z)}{dM} K^{y}_\ell(M,z) b(M,z)\,.
\label{eq:b_ell_y}
\ee

\subsection{Galaxy overdensity}\label{sub_sec:galaxy}
We model the distributions of galaxies inside dark matter halos using the halo occupation distribution (HOD) model. The mean occupation function of the central and satellite galaxies in a halo of mass $M_h$ are given by \citep{Zheng2004}
\be
\langle N_{\rm cen}(M_h) \rangle=\frac{1}{2}\left[ 1+ \mathrm{erf}\left(\frac{\log M_{\rm h}-\log M_{\rm th}}{\sigma_{\rm logM}}\right)\right]\,,
\label{eq:N_cen}
\ee
and
\be
\langle N_{\rm sat}(M_h) \rangle=\left( \frac{M_{\rm h}-M_{\rm cut}}{M_1}\right)^{\alpha_{\rm g}}\,.
\label{eq:N_sat}
\ee
Here, $\langle N_{\rm cen}(M_h) \rangle$ and $\langle N_{\rm sat}(M_h) \rangle$ are the average number density of central and satellite galaxies. $M_{\rm th}$ is the threshold halo mass to host a central galaxy, $M_{\rm cut}$ is the minimum mass for hosting satellite galaxies, $\sigma_{\rm logM}$ is the width of the transition of step-like error function, $\alpha_{\rm g}$ is the power law exponent, and $M_1$ is the mass normalization factor. The value of HOD-model parameters are given as $\log M_{\rm th}=12.712$, $\sigma_{\rm logM}=0.287$, $\log M_{\rm cut}=12.95$, $\log M_{1}=13.62$, and $\alpha=0.98$ \citep{Zheng2004}.

The Fourier transform of the distribution of satellite galaxies is given by \cite[e.g.][]{Scoccimarro2000, Pandey:2019}:
\be
K^g_{\ell}= \frac{W_{\rm g}(z)}{\chi^2 \bar{n}_{\rm g}(z)} \times \sqrt{2N_{\rm cen}N_{\rm sat}u_{\rm sat} + N_{\rm cen}^2N_{\rm sat}^2u_{\rm sat}^2}\,, \label{eq:K_ell_g}
\ee
where, $\bar{n}_{\rm g}(z)$ is the mean galaxy number density and $W_{\rm g}(z)$ is the normalized redshift distribution of galaxies that can be expressed as $\frac{c}{H(z)}\frac{dn_{\rm g}}{dz}$. We assume that the distribution of satellite galaxies follows the underlying dark matter overdensity field. The term $u_{\rm sat}$, for the truncated NFW profile is given by \citep{NFW}
\be
u_{\rm sat}=\frac{\delta_c\rho(z)}{x(1+x)^2} \,, \label{eq:usat}
\ee
where $\rho(z)$ is the mean density of the Universe, and $\delta_c$ is the characteristic over density of halos.

Putting everything together, the effective bias of the galaxies becomes
\begin{multline}
b_{\ell}^{\rm g} (M, z)=  \frac{W_{\rm g}(z)}{\chi^2 \bar{n}_{\rm g}(z)}  \int_{M_{\rm min}}^{M_{\rm max}} dM\frac{dn (M,z)}{dM}\\ N_{\rm cen}(M)(1+N_{\rm sat}(M)) u_{\rm sat} b(M,z)\, .
\label{eq:b_ell_g}
\end{multline}
\subsection{Power spectrum}
The 1-halo contribution to the power spectra can be expressed within the mass range, $M_{\rm min}$ to $M_{\rm max}$, and the redshift range between $z_{\rm min}$ and $z_{\rm max}$ as \citep{Komatsu1999, Cooray:2002dia}
\be
C_\ell^{{\rm XY}\,{\rm (1-halo)}} = \int_{z_{\rm min}}^{z_{\rm max}} dz \frac{d^2V}{d\Omega dz} \int_{M_{\rm min}}^{M_{\rm max}} dM\frac{dn (M,z)}{dM} K_\ell^{X} K_\ell^{Y}\,.\label{eq:cl1h}
\ee
Here, $K_\ell^{X}$ and $K_\ell^{Y}$ are the Fourier transform of \textit{kernel} of the X and Y observables, respectively. $dV$ is the comoving volume element at $\chi(z)$. 

The 2-halo term can be written as \citep{Komatsu1999, Cooray:2002dia}
\begin{multline}
C_\ell^{{\rm XY}\,{\rm (2-halo)}} = \int_{z_{\rm min}}^{z_{\rm max}} dz \frac{d^2V}{ d\Omega dz} b_\ell^{X}(M, z) b_{\ell}^{Y}(M,z) \\ \,P_{\rm lin} \left(k=\ell/\chi, k \right)\,,\label{eq:cl2h}
\end{multline}
where $P_{\rm lin}$ is the matter power spectrum that we calculated using CAMB\footnote{\href{https://camb.info}{https://camb.info}} in the linear regime \citep{Lewis:1999bs}. 
The cross power spectrum is the sum of 1-halo and 2-halo terms:
\be
C_\ell^{\rm XY} =C_\ell^{{\rm XY}\,{\rm (1-halo)}} +C_\ell^{{\rm XY}\,{\rm (2-halo)}}\,. \label{eq:cltot}
\ee 

Using Eq.\,\ref{eq:cl1h}, \,\ref{eq:cl2h} and \,\ref{eq:cltot}, we calculate the auto power spectrum by setting $X=Y$.
\section{Comparison with Simulations}\label{sec:simulation}
\begin{figure}[h]
 \includegraphics[width=0.47\textwidth]{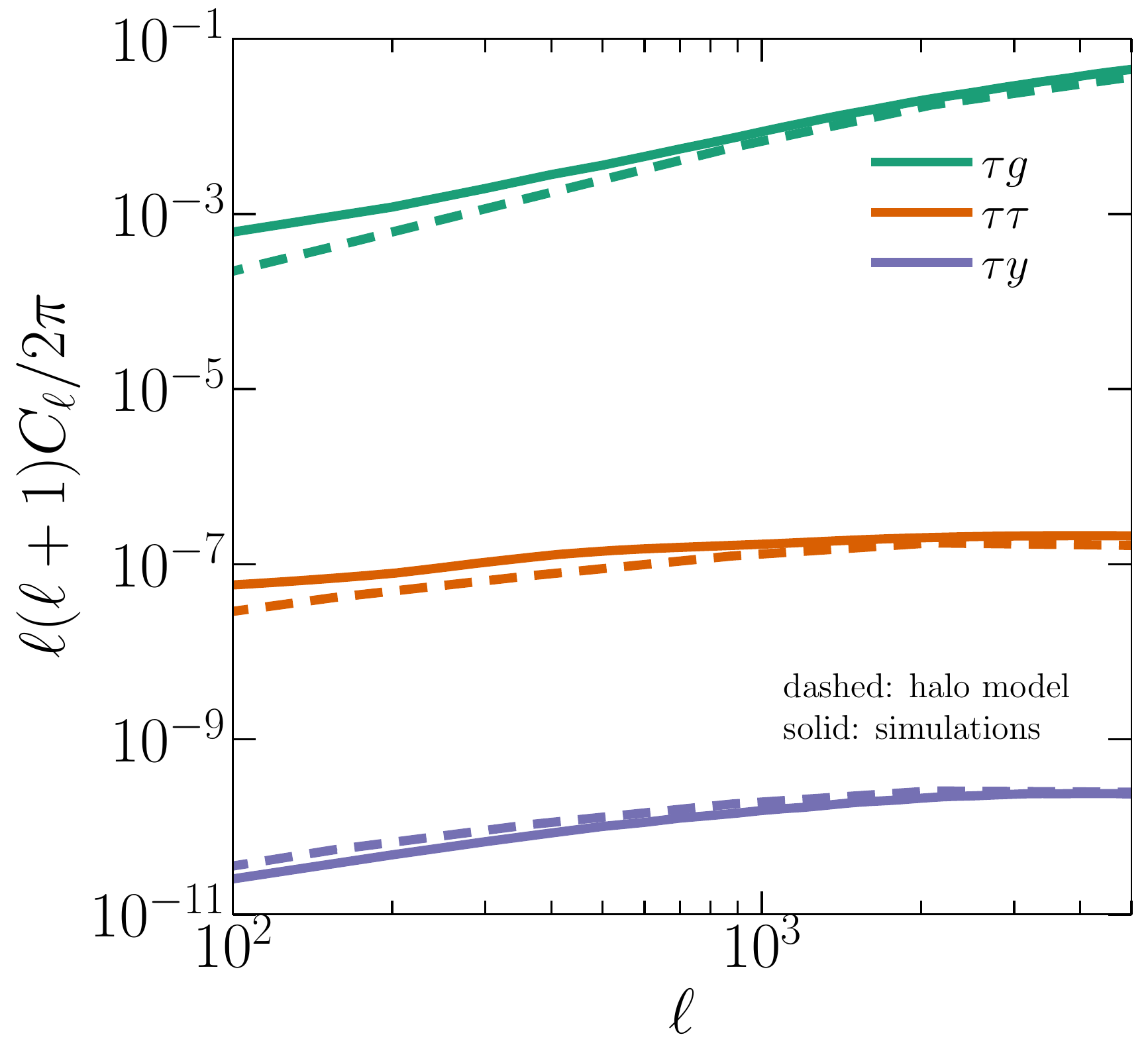}
  \caption{The agreement between the power spectra of $\tau g$, $\tau y$, and $\tau\tau$ calculated from the hydrodynamic simulations and analytic halo model. We fix $M_{\rm min}=3\times 10^{13}\, M_\odot$, $z_{\rm min}=0$ and $z_{\rm max}=5$. This validity check ensures us using halo model calculations reliably for constraining astrophysical parameters of gas thermodynamics in halos.}
  \label{fig:sims_comparison}
\end{figure}

We validate our results from the analytic halo model against existing cosmological hydrodynamic simulations. These simulations were performed using a modified version of the smoothed particle hydrodynamics (SPH) code, GADGET-2 \citep{Gadget}. Included in these simulations were sub-grid models for active galactic nuclei (AGN) feedback \citep[for more details see][]{BBPSS}, radiative cooling, star formation, galactic winds, supernova feedback \citep[for more details see][]{SpHr2003}, and cosmic ray physics \citep[for more details see][]{2006MNRAS.367..113P,2007A&A...473...41E,2008A&A...481...33J}. The box sizes for these simulations were 165 Mpc$/h$, with a resolution of 256$^3$ gas and dark matter (DM) particles. This yields mass resolutions of $M_\rmn{gas} = 3.2\times 10^{9} \rmn{M}_{\odot}/h$ and $M_\rmn{DM} = 1.54\times 10^{10} \rmn{M}_{\odot} /h$. Thus, these simulations can resolve more than 1000 particles in halos with masses of $M > 3\times 10^{13}\, M_\odot$. The cosmological parameters used for these simulations were $\Omega_\rmn{M} = \Omega_\rmn{DM} + \Omega_\rmn{b} = 0.25$, $\Omega_\rmn{b} = 0.043$, $\Omega_\Lambda = 0.75$, $H_0=100\,h\,\rmn{km}\,\rmn{s}^{-1}\,\rmn{Mpc}^{-1}$, $h=0.72$, $n_\rmn{s} =0.96$ and $\sigma_8 = 0.8$. These parameters differ from those used in the halo model but not at the level to cause substantial differences larger than what is currently shown in Figure \ref{fig:sims_comparison}.

We compare the amplitude of $C_{\ell}^{\tau g}$, $C_{\ell}^{\tau y}$, and $C_{\ell}^{\tau \tau}$ estimated from the halo-model against the hydrodynamic simulations. The halo-model approach is computationally efficient, and hence, if our results from the halo-model are consistent with the results from simulations, we can calibrate our model to study CGM. In addition, we have more freedom to select the different combinations of astrophysical parameters that we can constrain with future datasets. 

In Figure \ref{fig:sims_comparison} we show a comparison between the power spectrum estimated from the halo-model approach and hydrodynamic simulations. In doing so, we fixed the minimum halo mass to $3\times 10^{13}\,M_\odot$ so that it matches with the mass resolution of the simulations. At small scales, $\ell> 1000$, all power spectra agree very well. This consistency check allows us to study the detectability of this signal by changing the different values of $M_{\rm min}$.

\section{Noise Model}\label{sec:noise_model}
In this section, we review the reconstruction method for optical depth fluctuations and sources of noise for the measurement of the galaxy and tSZ power spectrum.

\subsection{$\tau$ reconstruction noise}
Future CMB experiments promise to reach high-resolution and low-noise measurements of polarization anisotropies, and it increases our ability to measure the $C_{\ell}^{\tau\tau}$ by applying a reconstruction method \citep{Roy:2018, Namikawa:2021zhh}. 
The fluctuations in optical depth introduce correlations between different Fourier modes. We can generalize the mode coupling of CMB temperature and polarization as \citep{Dvorkin:2008:tau-est}:
\be
\langle a^{X1}_{\ell_1 m_1} a^{X2}_{\ell_2 m_2}\rangle = \sum_{\ell m} \Gamma^{X_1 X_2}_{\ell_1 \ell_2 \ell} \begin{pmatrix} \ell_1 & \ell_2 & \ell\\ m_1 & m_2 & m \end{pmatrix} (\Delta \tau)^{*}_{\ell m}\,, \label{eq:mode_coupling}
\ee
where $X_1$ and $X_2$ are any combinations of $T$, $E$, and $B$, $\ell=\ell_1 + \ell_2$, and $\Gamma^{X_1 X_2}_{\ell_1 \ell_2 \ell}$ is the coupling factor that correlates $\ell_1$ and $\ell_2$. To minimize the noise of the estimator, we only consider the $EB$ quadratic estimator (QE) as it provides the highest signal-to-noise ratio \citep{Gluscevic:2012:tau}. The coupling factor for the $EB$ estimator can be written in terms of Wigner-3j symbols as
\begin{multline}
\Gamma^{EB}_{\ell_1 \ell_2 \ell}=\frac{C_{\ell_1}^{EE}}{2i} \sqrt{ \frac{(2\ell_1 + 1)(2\ell_2 + 1)(2\ell + 1)}{4\pi} }\\
\left[ \begin{pmatrix} \ell_1 & \ell_2 & \ell\\ -2 & 2 & 0 \end{pmatrix} - \begin{pmatrix} \ell_1 & \ell_2 & \ell\\ 2 & -2 & 0 \end{pmatrix} \right] \,,
\end{multline}
where $C_{\ell}^{EE}$ is the power spectrum of the $E$-mode polarization field.  
The noise for $\tau$ reconstruction depends on the sensitivity and resolution of a CMB experiment. The reconstructed $\tau$ field is given by \citep{Dvorkin:2008:tau-est}
\begin{multline}
\widetilde{\tau}^{*}_{\ell m}= N_{\ell}^{\tau \tau} \sum_{\ell_1 \ell_2} \sum_{m_1 m_2} \Gamma^{EB^*}_{\ell_1 \ell_2 \ell} \begin{pmatrix} \ell_1 & \ell_2 & \ell\\ m_1 & m_2 & m \end{pmatrix} \\ \times \frac{a^{E^*}_{\ell_1 m_1} a^{B^*}_{\ell_2 m_2}}{(C^{EE}_{\ell_1}+N^{EE}_{\ell_1})(C^{BB}_{\ell_2}+N^{BB}_{\ell_2})} \,, \label{eq:}
\end{multline}
where $C_{\ell}^{EE}$ and $C_{\ell}^{BB}$ are the power spectra of $E$ and $B$ modes.  We note that here we only consider the "screening" contribution arising from an inhomogeneous optical depth, and do not include the extra information from new polarization generated from scattering from incident quadrupoles \citep[e.g.,][]{Dvorkin:2009ah}.  For $C_{\ell}^{BB}$ our forecast includes iterated delensing \citep{Smith:2010gu} which we assume does not get modulated by the $B$ modes induced by $\tau$ fluctuations. $N_{\ell}^{EE}$ and $N_{\ell}^{BB}$ are instrumental noise power spectra for $E$ and $B$-mode signals, respectively. The instrumental noise power spectra are given by \citep{Knox:1995dq}
\begin{equation}
    N_{\ell}^{EE} = N_{\ell}^{BB} = \Delta_{\rm P}^2 \exp\left[\frac{ \ell(\ell+1)\theta_{\rm f}^2}{8\ln 2} \right]\,. \label{eq:nell_p}
\end{equation}
In the above equation, $\Delta_{\rm P}$ is the polarization sensitivity of a CMB experiment, which is $\sqrt 2$ times larger than the sensitivity in temperature, $\Delta_{\rm T}$, and $\theta_{\rm f}$ is the full width at half maximum (FWHM) of the beam size.

Finally, the reconstruction noise, $N_{\ell}^{\tau\tau}$ is given by \citep{Dvorkin:2008:tau-est}
\begin{multline}
    N_{\ell}^{\tau\tau} = \left[\frac{1}{2\ell + 1} \sum_{\ell_1 \ell_1}\frac{\left|\Gamma^{EB^*}_{\ell_1 \ell_2 \ell}\right|^2} {(C^{EE}_{\ell_1}+N^{EE}_{\ell_1})(C^{BB}_{\ell_2}+N^{BB}_{\ell_2})}\right]^{-1}\,. \label{eq:Nell_tau}
\end{multline}
We calculate the $\tau$ reconstruction noise for CMB S4 like experiments with 1$\mu K$ arcmin sensitivity in temperature and 1 arcmin beam size \citep{CMBS4}. Unless mentioned, we use $l_{min}=30$ and $l_{max}=5000$ throught this work because a CMB-S4-like experiment will make observations at these scales. 

\begin{figure}[h]
  \begin{center}
    \includegraphics[width=0.45\textwidth]{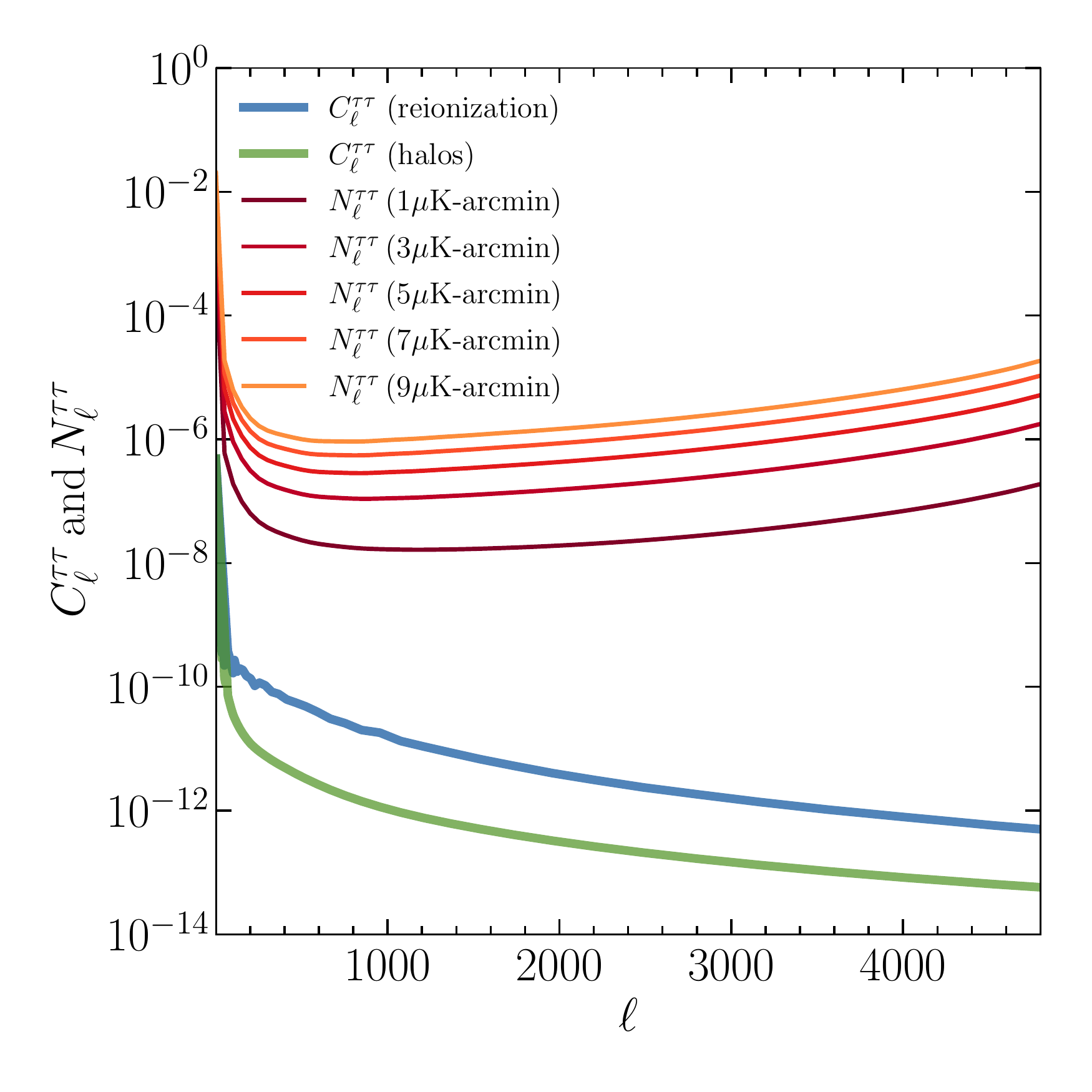}
    \caption{Reconstruction noise of optical depth power spectrum corresponding to different sensitivities in temperature of a CMB experiment while the FWHM of the beam is fixed at 1 arcmin. $C_\ell^{\tau\tau}$ from reionization and halos ($M>3 \times 10^{13}\,M_\odot$) are calculated from hydrodynamic simulations \citep{Battaglia2012}. }
    \label{fig:nltautau}
  \end{center}
\end{figure}

This process is analogous to the reconstruction method of the lensing potential from the CMB. In practice, both lensing and patchy reionization signals are present in the CMB data, and if we apply only the $\tau$ estimator, there will be leakage from the lensing field to the $\tau$ field \citep{Dvorkin:2008:tau-est, Su:2011ff}. This problem can be dealt with by applying a \enquote{bias-hardened} $\tau$ estimator that is insensitive to lensing at leading order fields \citep{Namikawa:2012:bhe,Namikawa:2021zhh}.   In order to determine the impact of bias-hardening the $\tau$ estimator from lensing, we computed the $\tau$ reconstruction noise $N_{\ell}^{\tau\tau}$  for bias-hardened estimators, for both $TT$ and $EB$.  We found that for the $TT$ estimator, the bias hardening can increase the lensing reconstruction noise $N_{\ell}^{\tau\tau}$ by factors of $2$--$3$ across all relevant $\ell$ values for the S4-like experiment. However, for the $EB$ estimator, bias hardening for lensing has a negligible effect on $N_{\ell}^{\tau\tau}$ for all but the very highest $\ell$ values, namely $\ell \gtrsim 4500$. Thus, we do not use the bias-hardened $EB$  noise curves in this work.

In Figure \ref{fig:nltautau}, we show the amplitude and shape of $N_\ell^{\tau\tau}$ that depends on the experiment. We vary the sensitivity of an experiment while keeping the beam size fixed to 1 arcmin. The reconstruction noise power spectrum decreases by a factor of $56$ if the sensitivity of a CMB experiment is decreased from 9 $\mu K$-arcmin to 1 $\mu K$-arcmin at $\ell=1000$. The power spectra of optical depth from reionization and halos at $\ell=1000$ are roughly 3 and 4 orders of magnitude smaller than the noise with $\Delta_T=1\,\mu K$-arcmin. 

\subsection{Noise for tSZ measurements}
We calculate the noise contribution in the $yy$ auto power spectrum, $N_{\ell}^{yy}$, by applying a constrained internal linear combination (cILC) algorithm. The cILC method depends on the assumption that the signal of our interest, tSZ, does not depend on the other signals and noise components and nulls a given contamination signal \citep{Eriksen:2004jg, Remazeilles:2010hq}. In the non-relativistic limit, the tSZ spectral dependence is very well defined, and it helps to minimize the variance to reconstruct the signal from the total observed signal. Since we are interested in measuring the $yy$ auto power spectrum in the absence of the cosmic infrared background (CIB, generally the largest contaminant to tSZ observations), we can write down a model for an angular scale ($d$) per frequency ($\nu$) as \citep{2016A&A...594A..25P}
\begin{equation}
    d_{\ell,\nu}=a_{\ell,\nu}f^{y}_{\nu}C_{\ell}^y+b_{\ell,\nu}f^{\mathrm{CIB}}_{\nu}C_{\ell}^\mathrm{CIB}+n_{\ell,\nu},
\end{equation}
where $a$ and $b$ are the scale and frequency-dependent coefficients. We will refer to $n$ as noise, but it also contains primordial CMB fluctuations, radio sources, instrumental noise, etc. The cILC algorithm constructs a weight vector $W$ such that we get a unit response to $a$ (the tSZ coefficients) and a null response to $b$ (the CIB coefficients), $W^T\vec{a}=1\quad W^T\vec{b}=0$. The general solution for $W$ was shown in \citet{Remazeilles:2010hq},
\begin{equation}
    W_{a=1,b=0}=\frac{b^Tn^{-1}ba^Tn^{-1}-a^Tn^{-1}bb^Tn^{-1}}{a^Tn^{-1}ab^Tn^{-1}b-(a^Tn^{-1}b)^2}.
\end{equation}
Note that we dropped the frequency dependence for clarity.
From the weight we can calculate $N_{\ell}^{yy}$ via
\begin{equation}
    N_{\ell}^{yy}=\sum_{\nu}(W_{\nu}n_{\nu}W_{\nu}^T)_{\ell}.
\end{equation}
We use the \textsc{SILC}\footnote{\href{https://github.com/nbatta/SILC}{https://github.com/nbatta/SILC}} package to estimate $N_{\ell}^{yy}$ in presence of primordial CMB, cosmic infrared background, foreground due to the dust and synchrotron radiation. In the near future, low noise experiments will be able to produce signal dominated measurements of the tSZ effect \citep{Hanany:2019lle}. 

\subsection{Shot noise of galaxy power spectrum}
For the case of the noise power spectrum of galaxies, we assume that the process of local galaxy formation in halos is linear and that the dominant source of noise is the shot noise term for the measurement of galaxy power spectra. In this regime, the shot noise becomes equal to $1/\bar{n}_{\rm g}$, where $\bar{n}_{\rm g}$ is the average number density of galaxies.
The average number density of galaxies at $z$ can be written as
\be
\bar{n}_g(z)=\int_{M_{ min}}^{M_{max}} dM \frac{dn(M,z)}{dM} N_g(M,z)\,. \label{eq:ngbar}
\ee

We consider the specifications of VRO\footnote{\href{https://www.lsst.org/}{https://www.lsst.org/}} to determine the shot noise term. We fix $M_{\rm max}=10^{13}\,M_\odot$ and consider a redshift distribution of galaxies in the redshift raange $z_{\rm min}=0$ and $z_{\rm max}=5$ as described in \citet{LSST-science-book}. 
The total number of galaxies that will be probed by a VRO-like-experiment is 50 per square arc-minute. Using Eq.\,\ref{eq:ngbar}, we find $M_{\min}=2\times 10^{11}\, M_\odot$ that satisfies this condition and we use this value of $M_{\rm min}$ for making forecasts for a VRO-like experiment. 

\section{Detectability of signals}\label{sec:result}
In this section, we discuss the detectability of the cross-correlation between different observational probes that are described in Section \ref{sec:halomodel}. For the forecasts, we used the sensitivity of CMB-S4 like experiment with the sensitivity, $\Delta_T= 1\,\mu K$-arcmin, and the FWHM of the beam, $\theta_{\rm f}=1$ arcmin.

As discussed in the previous section, we assume that the estimator described in \citet{Dvorkin:2008:tau-est} to reconstruct a fluctuating $\tau$ map, and this map is then cross-correlated with a given galaxy sample or with a Compton-$y$ map, which produces cross spectra ($C^{\tau g}_\ell$ and $C^{\tau y}_\ell$, respectively). A similar approach was proposed to measure the fluctuations in $\tau$ from reionization by cross-correlating with measurements from 21cm experiments \citep{Meerburg2013, Roy:2019qsl}.

The total $\rmn{S} / \rmn{N}$ is calculated by summing the signal-to-noise per $\ell$ mode,
\be
\left(\rmn{S} / \rmn{N}\right)^2 = \sum_\ell{\left(\frac{C_\ell^{XY}}{\Delta C_\ell^{XY} }\right)^2},
\ee \label{eq:snr_form}
where $\Delta C_\ell^{XY}$ is theoretical error on $C_\ell^{XY}$ given $N_\ell^{\tau\tau}$. The exact form of $\Delta C_\ell^{XY}$ we use is 
\begin{multline}
\left (\Delta C_\ell^{XY} \right)^2 = \frac{1}{(2\ell + 1)f^{XY}_\rmn{sky}} [(C_\ell^{XY})^2 + \\
\left(C_\ell^{XX}+N_\ell^{XX}\right)\left(C_\ell^{YY} + N_\ell^{YY} \right)]\,,
\label{eq:delcl}
\end{multline}
where $f^{XY}_\rmn{sky}$ is the fraction of sky covered, $C_\ell^{XX}$ and $N_\ell^{YY}$ are the auto spectrum and noise term of a given tracer, respectively. Here we assume that $N_\ell^{\tau\tau} \gg C_\ell^{\tau\tau}$ to simplify Eq.\, \ref{eq:delcl}. 
\begin{figure}[h]
  \includegraphics[width=0.45\textwidth]{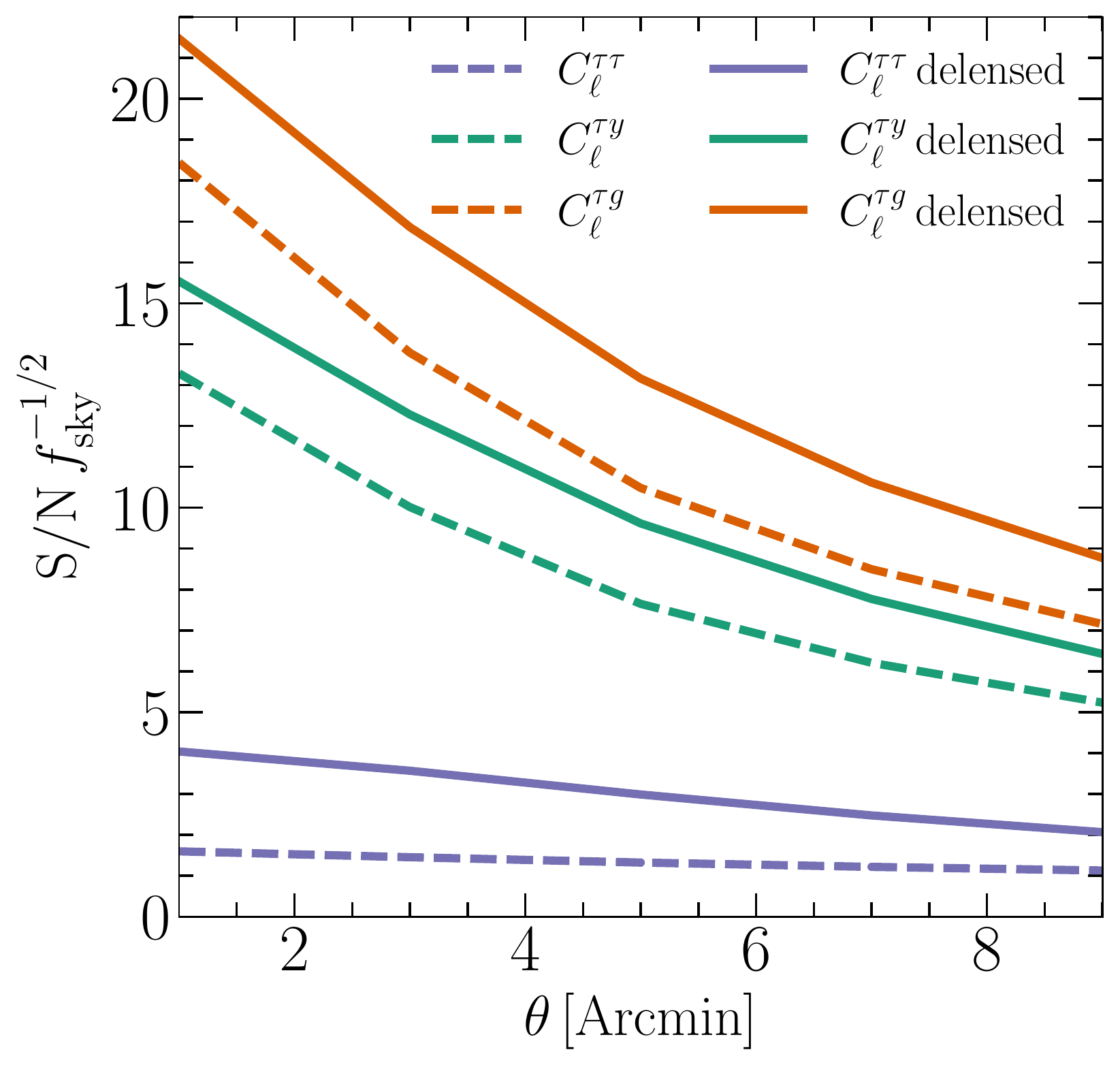}
  \caption{ Forecast of signal-to-noise ratio of $C_\ell^{\tau g}$, $C_\ell^{\tau y}$, and $C_\ell^{\tau \tau}$ signals with FWHM of beam for CMB-S4 like experiment for $M_{\rm min}=2\times 10^{11}\,M_\odot$ and $M_{\rm max}=10^{13}\,M_\odot$. The cross-correlations are calculated using analytic halo model and the instrumental sensitivity in temperature is set to 1 $\mu K$-arcmin. The signal-to-noise ratio increases with the smaller beam size because the number of accessible modes increases with resolutions of an experiment. In addition to that, delensing of $B$ modes help to reduce the noise in polarization that lower the reconstruction noise of $\tau$ fluctuations.}
  \label{fig:snr_sims}
\end{figure}

\begin{figure*}[t]
  \includegraphics[width=\textwidth]{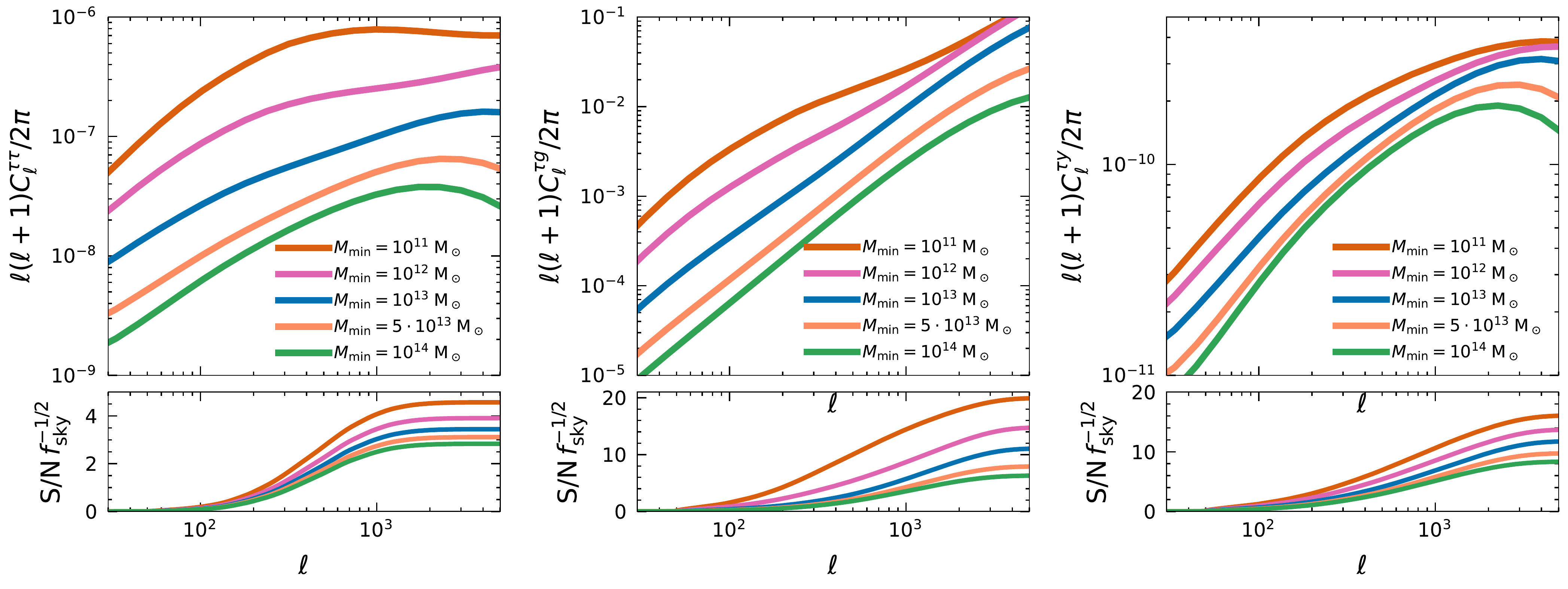}
  \caption{\textit{Top panel}: $\tau\tau$ (left panel), $\tau g$ (middle panel), and  $\tau y$ (right panel) power spectra shown as a function of the minimum halo mass, $M_{\rm min}$. \textit{Bottom panel}: cumulative SNR calculated for the $\tau\tau$ (left panel), $\tau y$ (middle panel), and $\tau g$ (right panel) with $1\,\mu K$arcmin sensitivities in temperature and 1\,arcmin beam size. This figure shows that signal-to-noise ratio are dominated by the low mass halos ($M_{\rm halo}\lesssim 10^{12}\,M_\odot$) rather than the high mass halos ($M_{\rm halo}\gtrsim 10^{13}\,M_\odot$).}
  \label{fig:all_signals}
\end{figure*}

\begin{figure*}[h]
\centering
  \includegraphics[width=0.6\textwidth]{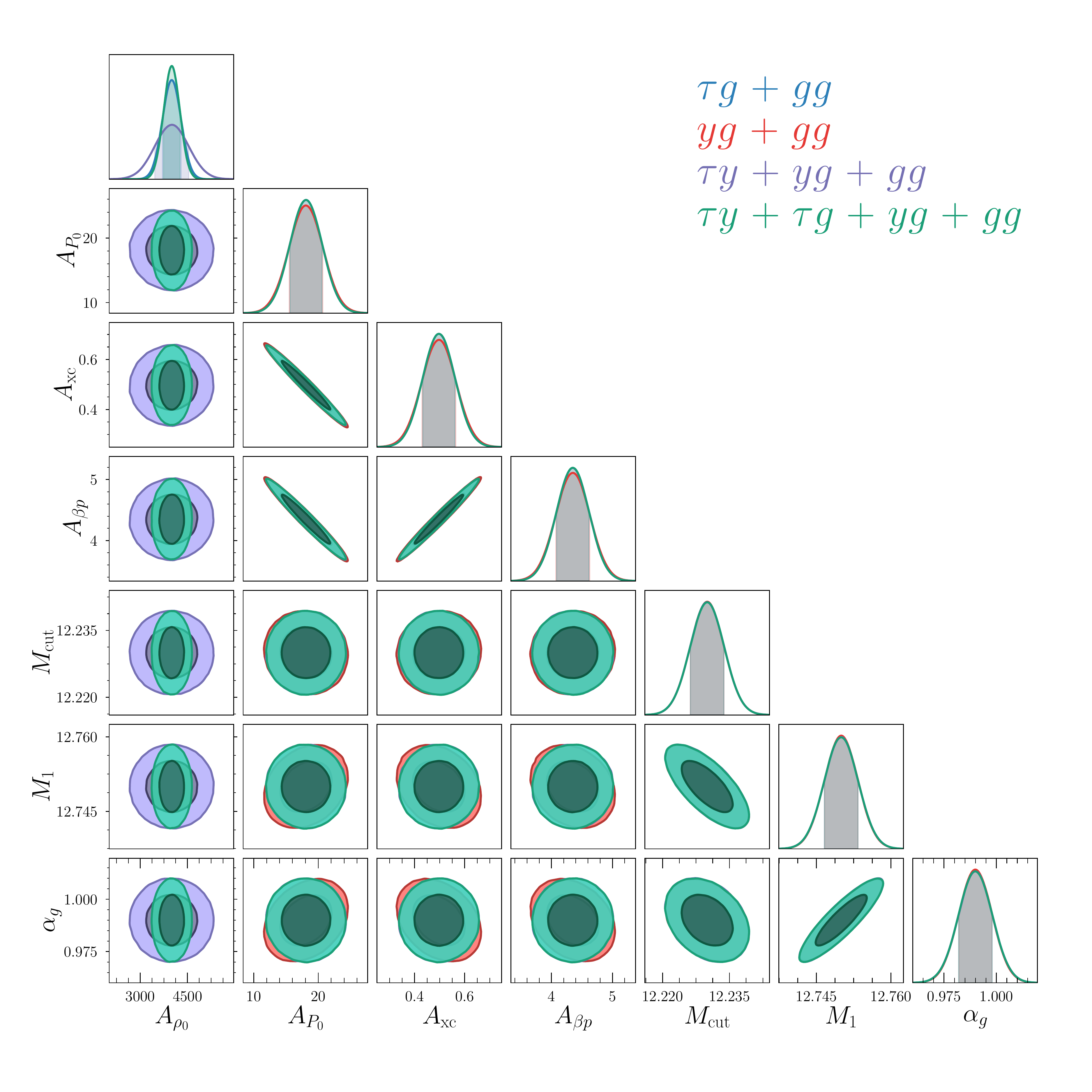}
  \caption{Forecasted constraints on the density, pressure and HOD parameters from the different combinations of $\tau g$, $\tau y$, and $yg$. We set $M_{\rm min}=2 \times 10^{11}\,M_\odot$ and $M_{\rm max}=10^{13}M_\odot$ to study the properties of circumgalactic medium. Tight constraints are placed on HOD parameters from the measuremnts of $C_{\ell}^{gg}$. }
  \label{fig:fisher_main}
\end{figure*}

\begin{figure*}[h]
  \centering
  \subfigure{\includegraphics[width=0.46\textwidth]{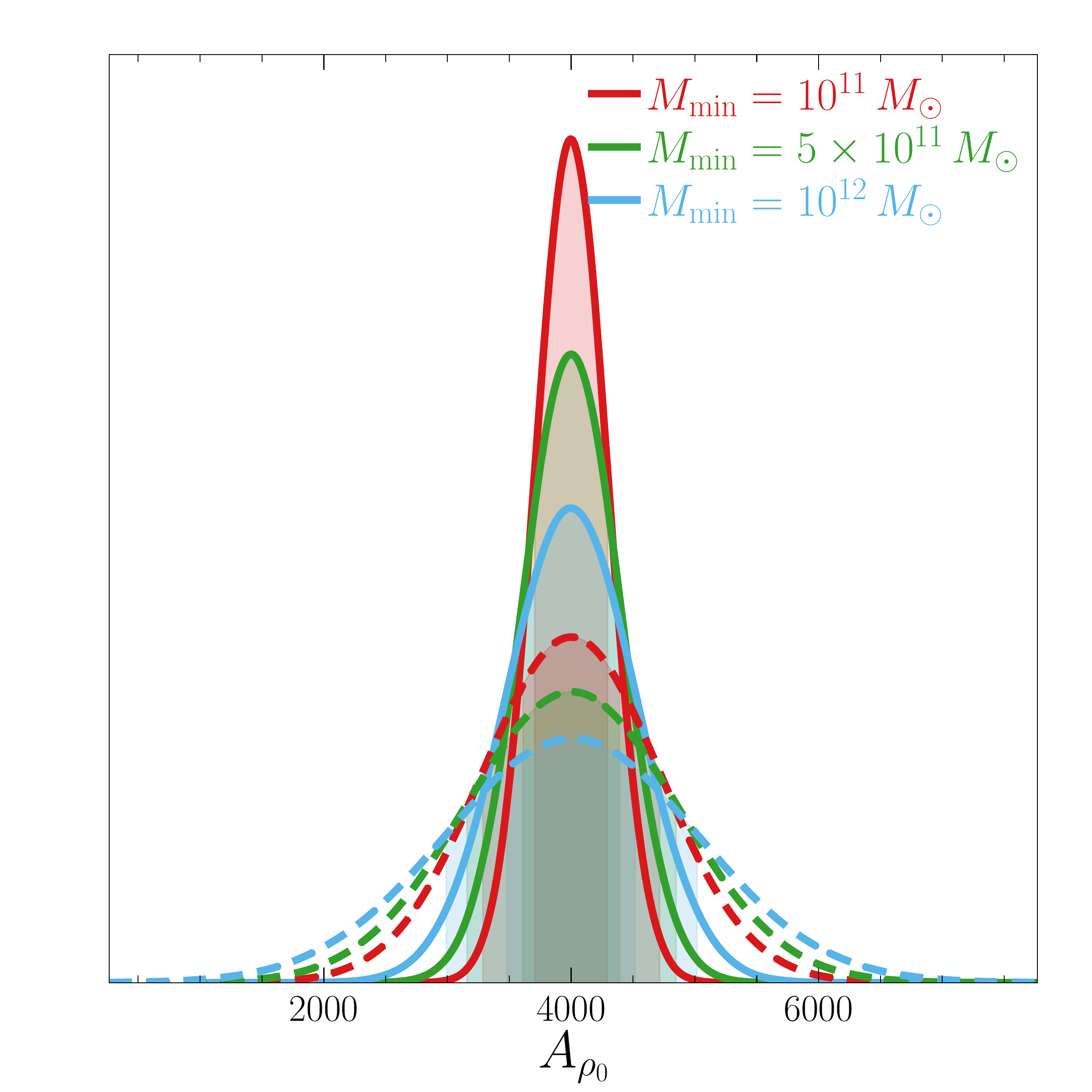}}\quad
  \subfigure{\includegraphics[width=0.45\textwidth]{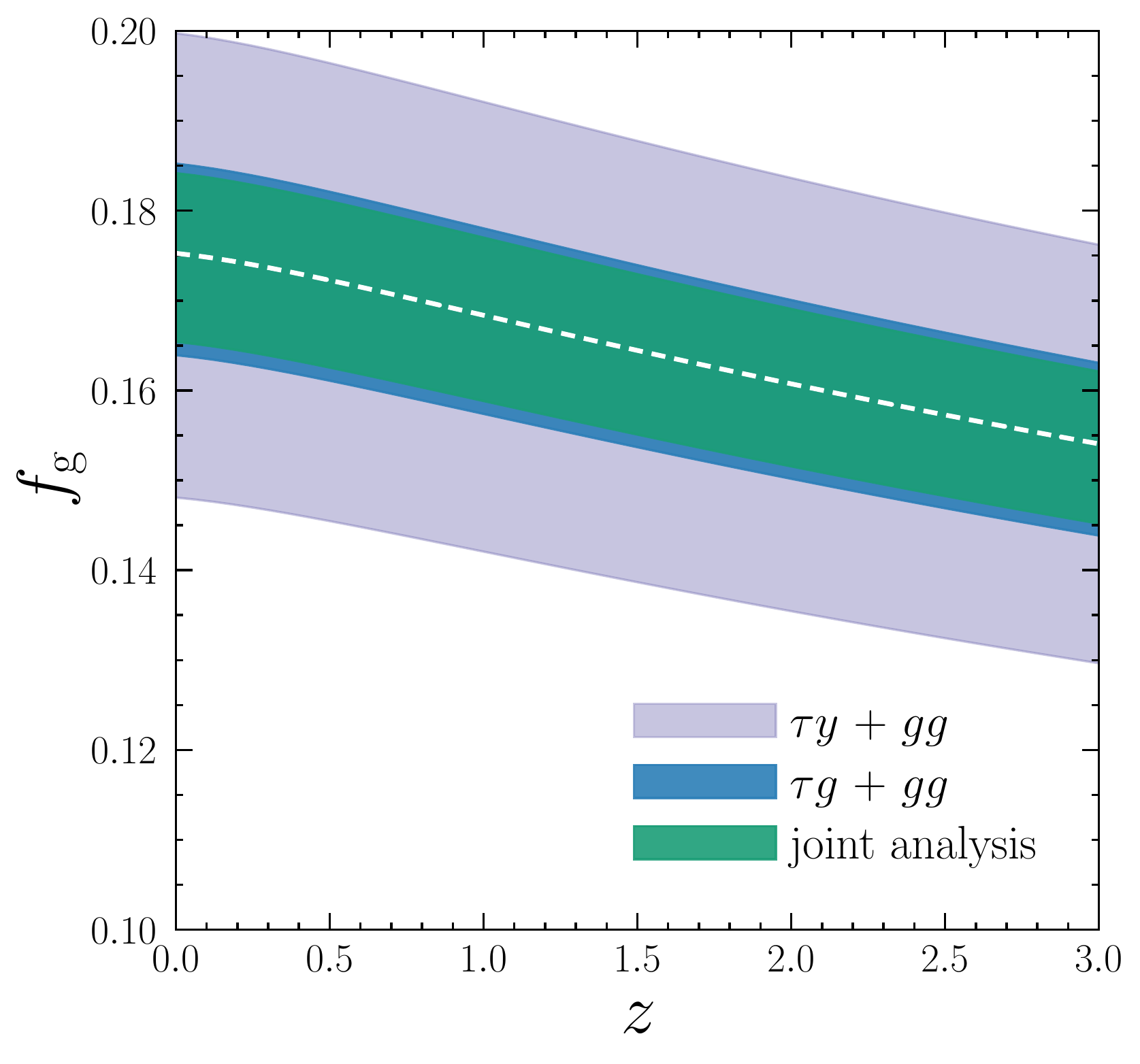}}
  \caption{\textit{left:} solid lines show the constraints on the amplitude of the density profile, $A_\rho$, from $\tau g$ cross-correlations and dashed lines represent the constraints inferred from $\tau y$ cross-correlation. \textit{right:} The forecasted 1$\sigma$ uncertainty of gas mass fraction, $f_{\rm g}$, calculated from a joint analysis of different probes of cross-correlation. Dashed white line is the best fit  curve for the $\tau y$ + $\tau g$ + $yg$ + $gg$.}
  \label{fig:rho_constraints}
\end{figure*}
We use the simulations described in Section \ref{sec:simulation} to calculate the cross power spectra $C^{\tau g}_\ell$ and $C^{\tau y}_\ell$. Here the field $g$ refers to a number-count map where a non-zero value is assigned at the location of each halo above a given minimum mass, and $y$ is a Compton-$y$ map with the same mass threshold. We cross-correlate the number-count and Compton-$y$ maps with the simulated $\tau$ maps. Then, for each redshift, we average over all simulation realizations and add these differential cross-spectra to get the signal.
\begin{table*}[t]
\setlength{\tabcolsep}{3pt}
\centering
\renewcommand{\arraystretch}{1.0}
\begin{tabular}{ccccccc}
\hline\hline
Parameters & meaning & fiducial values & $\tau g$ + $gg$ & $\tau y$ + $yg$+ $gg$ & $yg$ + $gg$ & joint analysis \\
\hline
\centering
$A_{\rho_0}$ & amplitude of density profile  & $4000$  & $7.27$  &$13.2$ & - & $6.38$ \\
$A_{P_0}$ & amplitude of pressure profile  & 18.1 & - & $13.64$ & $14.32$ & $13.65$ \\
$A_{\rm xc}$ & amplitude of characteristics scale of halo  & 0.497 & - & $12.9$ & $13.54$  & $12.86$\\
$A_{\beta p}$ & power law index of the fall of pressure profile & 4.35 & - & $6.0$  & $6.35$ & $6.0$\\
\hline
\end{tabular}
\caption{The forecasted $1\sigma$ uncertainties on the parameters of density and pressure profiles.}
\label{table:fisher}
\end{table*}

Note that all the cross-correlations have contributions both from the high redshift ($z\gtrsim5$) and low redshift ($z\lesssim 5$). The $yy$ from reionization is subdominant as the power spectrum is at least 3-4 orders of magnitude smaller than the galaxy and cluster contributions \citep{Hill:2015}, but the amplitude of the $\tau y$ power spectrum from reionization is comparable with the contributions from halos at low redshift (though they may have different shapes as the characteristic scale of ionized bubbles and halos are different; \cite{Hill:2015, Namikawa:2021zhh}). We calculate the $\tau y$ signal for a fiducial model of reionization with the characteristic bubble size of ionized bubbles $\bar{R}_b=5\,$ Mpc following a semi-analytic model described in \cite{Namikawa:2021zhh}. Hence, we consider the $\tau y$ signal from reionization as a source of noise in our forecasts. $yg$ and $\tau g$ from high and low redshift do not correlate as we choose the galaxy distribution function at the low redshift only, so they have different kernels. 

In Figure \ref{fig:snr_sims}, we show the total $\rmn{S} / \rmn{N}$ as a function of beam size for CMB-S4 like survey that has a noise level of 1$\mu$K acrmin. The $\ell$ range we consider is from 100-5000 in polarization. For a halo mass threshold of $M_{\rm min}=2\times 10^{11} M_\odot$ and $\theta_{\rm f}=1$ arcmin, we find the total $\rmn{S} / \rmn{N} f_\rmn{sky}^{-1/2} \approx 15$ and 10 for $C^{\tau g}_\ell$ and $C^{\tau y}_\ell$, respectively. After delensing the lensed $B$-mode power spectrum at level of 90\% \citep{Smith:2010gu, CMBS4}, we find that the signal-to-noise ratio increases by 14\% and 12.5\% for $C^{\tau g}_\ell$ and $C^{\tau y}_\ell$, respectively. 

In Figure \ref{fig:all_signals}, we show the variation of $\tau y$, $\tau g$, and $\tau y$ signals for different minimum halo mass, $M_{\rm min}$. 
The cumulative SNRs for $\tau g$, $\tau y$ and $\tau\tau$ are 20, 16 and 5.5 corresponding to $M_{\rm min}=10^{11}\,M_\odot$. The SNRs change to 11, 11.5 and 4 if $M_{\rm min}$ is set to $10^{13}\, M_\odot$. We find SNR for $yg$ is 1710 and 950 for $M_{\rm min}= 10^{11}\, M_\odot$ and $10^{14}\, M_\odot$ respectively. 

\section{Fisher Forecast}\label{sec:fisher}
In this section, we discuss how the future measurements of these cross-correlations can be used to constrain the astrophysical parameters related to the pressure and density of CGM. For the rest of our analysis, we use the upper limit on the halo mass of $M_{\rm max}=10^{13}\, M_\odot$ so that we can study the properties of CGM. We assume that the pressure and density profile is self-similar to the redshift and mass of halos.

We use Fisher matrix analysis to forecast the measurement uncertainties of the parameters of our interest. The Fisher matrix is given by

\be
F_{ij}=\sum_{\ell} \frac{1}{(\Delta C^{XY}_\ell)^2} \frac{\partial C_\ell^{XY}}{\partial P_i}\frac{\partial C_\ell^{XY}}{\partial P_j} \,,
\label{eq:fij}
\ee 
where $P_i$ and $P_j$ are the parameters which we aim to constrain from the measurements of cross-correlations. 

We select parameters of pressure and density as described in Table \ref{table:fisher} and show the constraints on parameters in Figure \ref{fig:fisher_main} for the different combinations of the $\tau g$, $yg$, $\tau y$, and $gg$ signal. Note that the $yg$ signal can probe the parameters related to the pressure profile, whereas the $\tau g$ signal is sensitive to the density profile of halos. The $\tau y$ signal is sensitive to both the pressure and density profile, and for this reason, we use it to break the degeneracy between the amplitude of density and pressure profiles. We use $M_{\rm min}= 2 \times 10^{11}\, M_\odot$, $z_{max}=5$, and $f_{\rm sky}= 0.5$ for this analysis. The Galaxy power spectrum puts tighter constraints on the HOD model than the parameters that are inferred from the $yg$ signal, and it helps to break the degeneracy between parameters of pressure and the HOD model. We present our forecasts of the parameters in Table \ref{table:fisher}. 

The detection of these cross-correlations leads us to understand some other astrophysical parameters that are directly related to the pressure and density profiles. We specifically focus on the gas mass fraction, $f_{\rm g}={M_{\rm gas}}/M_{\rm halo}$. It should be noted that this quantity depends on the definition of radius and mass of halos. For example, the value of $f_{\rm g}$ will be different for radius $R_{200}$, $R_{500}$, and $R_{2500}$ as they correspond to different halo masses $M_{200}$, $M_{500}$, and $M_{2500}$. 

The $M_{\rm gas}$ inside $R_{500}$ can be written as
\be
M_{\rm gas, 500}(z)=\int_{0}^{R_{500}}\,\rho_{\rm gas}(r,z)\;4\pi r^2 dr\,.
\label{eq:mgas}
\ee
In Figure \ref{fig:rho_constraints}, we show the forecasted constraints on the amplitude of the density profile $A_\rho$ from the $\tau g$ and $\tau y$ signals. As expected, $\tau g$ put tighter limits on the $A_\rho$ as the SNR is larger than the $\tau y$ for the same $M_{\rm min}$. We find that the gas mass fraction, $f_{g\rm }$, can be measured with more than $7\sigma$ confidence for $\tau y$ + $gg$ and $17\sigma$ for $\tau g$ + $gg$. 
\section{Discussion}\label{sec:discussion}
In the near future, there will be signal dominated high-resolution tSZ measurements that will bring a unique opportunity to break the degeneracy among astrophysical parameters related to the thermodynamic properties of CGM by performing cross-correlation with other probes. We explored such cross-correlations to measure the density and pressure profiles of halos by independent observational probes. In this paper, we presented the
detectability of cross-correlations between the optical depth anisotropy (reconstructed from the CMB polarization anisotropy), Compton-y parameter (derived from the CMB temperature measurements), and galaxy overdensity field. The optical depth anisotropy is a tracer of the gas density distribution in halos, the Compton $y$ parameter probes the pressure profile, and the galaxy number-count carries information of the distribution of galaxies in dark matter halos. 

We used a semi-analytic halo-model formalism (compared against hydrodynamic simulations) to estimate the power spectra of these cross-correlated signals. The $yg$ signal is the most promising probe for constraining the pressure parameters, which can be performed using VRO galaxies and Compton $y$ map from the CMB-S4 experiment. We find the signal-to-noise ratio can reach as high as 500 to 2000 depending on the $M_{\rm min}$. Such a high signal-to-noise ratio can be used to understand the sub-grid physics in halos, such as AGN feedback, turbulence, and depletion efficiency of gas in CGM. 

On the other hand, the $\tau g$ signal is a useful probe for measuring the density profile along with the priors on HOD model parameters that can be well achieved by the precise measurement of the $gg$ signal alone. The measurements of pressure and density profiles by different probes like $\tau g$, $yg$, and $yg$ will also lead us to determine the temperature profile of halos. A joint analysis will not only improve our understanding of the low redshift Universe but will also help us to investigate the reionization process. The $y\tau$ signal at low redshift contaminates the $y\tau$ signal from reionization. Hence a proper estimation of $y\tau$ from halos will help to remove it from the total observed signal that can put tighter constraints on reionization parameters such as size and temperature of ionized bubbles \citep{Namikawa:2021zhh}.

The multiple probes presented in this paper are complementary to tSZ and kSZ measurements and their cross-correlations with galaxy surveys. Previous kSZ measurements have probed the average density profile of halos \citep{2016PhRvD..93h2002S, AtacamaCosmologyTelescope:2020wtv, Amodeo:2020mmu}. However, spectroscopic galaxy redshifts are required to reconstruct the velocity field to achieve these constraints on the density profile. The cross-correlations we propose do not require spectroscopic information of galaxies. Instead, we forecast complementary and comparable constraints by utilizing a larger sample of galaxies selected from future imaging surveys. Our forecasts show that the cross-correlation between $\tau$ and $g$ is detectable with more than 15 $\sigma$ for upcoming experiments, whereas the auto power spectrum of $\tau$ is detectable at $\sim 4 \sigma$.

Baryons that are outside the halos are hard to see in optical and X-ray bands, but their presence is encoded in the optical depth anisotropies as these baryons interact with CMB photons via Thomson scattering. The cross-correlations presented here are sensitive to the global properties of gas inside and outside halos. Thus, probing the baryon abundances inside and outside halos with such cross-correlations can be used to resolve the question of are we missing baryons? Furthermore, we demonstrated that our proposed joint cross-correlations are robust. We found that \enquote{bias hardening} the estimator to reduce leakage from CMB lensing did not have an appreciable impact on the results.

\begin{acknowledgments}
The authors would like to thank Stefania Amodeo, Joel Meyers, Connor Sheere, Kendrick Smith, and David Spergel. VG is supported by the National Science Foundation under Grant No. PHY-2013951. NB acknowledges support from NSF grant AST-1910021 and NASA grants 21-ADAP21-0114 and 21-ATP21-0129.
\end{acknowledgments}
\bibliography{mybib}{}

\begin{thebibliography}{}
\expandafter\ifx\csname natexlab\endcsname\relax\def\natexlab#1{#1}\fi
\providecommand{\url}[1]{\href{#1}{#1}}
\providecommand{\dodoi}[1]{doi:~\href{http://doi.org/#1}{\nolinkurl{#1}}}
\providecommand{\doeprint}[1]{\href{http://ascl.net/#1}{\nolinkurl{http://ascl.net/#1}}}
\providecommand{\doarXiv}[1]{\href{https://arxiv.org/abs/#1}{\nolinkurl{https://arxiv.org/abs/#1}}}

\bibitem[{{Abazajian} {et~al.}(2019){Abazajian}, {Addison}, {Adshead}, {Ahmed},
  {Allen}, {Alonso}, {Alvarez}, {Anderson}, {Arnold}, {Baccigalupi}, {Bailey},
  {Barkats}, {Barron}, {Barry}, {Bartlett}, {Basu Thakur}, {Battaglia},
  {Baxter}, {Bean}, {Bebek}, {Bender}, {Benson}, {Berger}, {Bhimani},
  {Bischoff}, {Bleem}, {Bocquet}, {Boddy}, {Bonato}, {Bond}, {Borrill},
  {Bouchet}, {Brown}, {Bryan}, {Burkhart}, {Buza}, {Byrum}, {Calabrese},
  {Calafut}, {Caldwell}, {Carlstrom}, {Carron}, {Cecil}, {Challinor}, {Chang},
  {Chinone}, {Cho}, {Cooray}, {Crawford}, {Crites}, {Cukierman}, {Cyr-Racine},
  {de Haan}, {de Zotti}, {Delabrouille}, {Demarteau}, {Devlin}, {Di Valentino},
  {Dobbs}, {Duff}, {Duivenvoorden}, {Dvorkin}, {Edwards}, {Eimer}, {Errard},
  {Essinger-Hileman}, {Fabbian}, {Feng}, {Ferraro}, {Filippini}, {Flauger},
  {Flaugher}, {Fraisse}, {Frolov}, {Galitzki}, {Galli}, {Ganga}, {Gerbino},
  {Gilchriese}, {Gluscevic}, {Green}, {Grin}, {Grohs}, {Gualtieri}, {Guarino},
  {Gudmundsson}, {Habib}, {Haller}, {Halpern}, {Halverson}, {Hanany},
  {Harrington}, {Hasegawa}, {Hasselfield}, {Hazumi}, {Heitmann}, {Henderson},
  {Henning}, {Hill}, {Hlozek}, {Holder}, {Holzapfel}, {Hubmayr},
  {Huffenberger}, {Huffer}, {Hui}, {Irwin}, {Johnson}, {Johnstone}, {Jones},
  {Karkare}, {Katayama}, {Kerby}, {Kernovsky}, {Keskitalo}, {Kisner}, {Knox},
  {Kosowsky}, {Kovac}, {Kovetz}, {Kuhlmann}, {Kuo}, {Kurita}, {Kusaka},
  {Lahteenmaki}, {Lawrence}, {Lee}, {Lewis}, {Li}, {Linder}, {Loverde},
  {Lowitz}, {Madhavacheril}, {Mantz}, {Matsuda}, {Mauskopf}, {McMahon},
  {McQuinn}, {Meerburg}, {Melin}, {Meyers}, {Millea}, {Mohr}, {Moncelsi},
  {Mroczkowski}, {Mukherjee}, {M{\"u}nchmeyer}, {Nagai}, {Nagy}, {Namikawa},
  {Nati}, {Natoli}, {Negrello}, {Newburgh}, {Niemack}, {Nishino}, {Nordby},
  {Novosad}, {O'Connor}, {Obied}, {Padin}, {Pandey}, {Partridge}, {Pierpaoli},
  {Pogosian}, {Pryke}, {Puglisi}, {Racine}, {Raghunathan}, {Rahlin},
  {Rajagopalan}, {Raveri}, {Reichanadter}, {Reichardt}, {Remazeilles}, {Rocha},
  {Roe}, {Roy}, {Ruhl}, {Salatino}, {Saliwanchik}, {Schaan}, {Schillaci},
  {Schmittfull}, {Scott}, {Sehgal}, {Shandera}, {Sheehy}, {Sherwin},
  {Shirokoff}, {Simon}, {Slosar}, {Somerville}, {Spergel}, {Staggs}, {Stark},
  {Stompor}, {Story}, {Stoughton}, {Suzuki}, {Tajima}, {Teply}, {Thompson},
  {Timbie}, {Tomasi}, {Treu}, {Tristram}, {Tucker}, {Umilt{\`a}}, {van
  Engelen}, {Vieira}, {Vieregg}, {Vogelsberger}, {Wang}, {Watson}, {White},
  {Whitehorn}, {Wollack}, {Kimmy Wu}, {Xu}, {Yasini}, {Yeck}, {Yoon}, {Young},
  \& {Zonca}}]{CMBS4}
{Abazajian}, K., {Addison}, G., {Adshead}, P., {et~al.} 2019, arXiv e-prints,
  arXiv:1907.04473.
\newblock \doarXiv{1907.04473}

\bibitem[{{Abell} {et~al.}(2009){Abell}, {Allison}, {Anderson}, {Andrew},
  {Angel}, {Armus}, {Arnett}, {Asztalos}, {Axelrod}, {Bailey}, {Ballantyne},
  {Bankert}, {Barkhouse}, {Barr}, {Barrientos}, {Barth}, {Bartlett}, {Becker},
  {Becla}, {Beers}, {Bernstein}, {Biswas}, {Blanton}, {Bloom}, {Bochanski},
  {Boeshaar}, {Borne}, {Bradac}, {Brandt}, {Bridge}, {Brown}, {Brunner},
  {Bullock}, {Burgasser}, {Burge}, {Burke}, {Cargile}, {Chandrasekharan},
  {Chartas}, {Chesley}, {Chu}, {Cinabro}, {Claire}, {Claver}, {Clowe},
  {Connolly}, {Cook}, {Cooke}, {Cooray}, {Covey}, {Culliton}, {de Jong}, {de
  Vries}, {Debattista}, {Delgado}, {Dell'Antonio}, {Dhital}, {Di Stefano},
  {Dickinson}, {Dilday}, {Djorgovski}, {Dobler}, {Donalek}, {Dubois-Felsmann},
  {Durech}, {Eliasdottir}, {Eracleous}, {Eyer}, {Falco}, {Fan}, {Fassnacht},
  {Ferguson}, {Fernandez}, {Fields}, {Finkbeiner}, {Figueroa}, {Fox},
  {Francke}, {Frank}, {Frieman}, {Fromenteau}, {Furqan}, {Galaz}, {Gal-Yam},
  {Garnavich}, {Gawiser}, {Geary}, {Gee}, {Gibson}, {Gilmore}, {Grace},
  {Green}, {Gressler}, {Grillmair}, {Habib}, {Haggerty}, {Hamuy}, {Harris},
  {Hawley}, {Heavens}, {Hebb}, {Henry}, {Hileman}, {Hilton}, {Hoadley},
  {Holberg}, {Holman}, {Howell}, {Infante}, {Ivezic}, {Jacoby}, {Jain}, {R},
  {Jedicke}, {Jee}, {Garrett Jernigan}, {Jha}, {Johnston}, {Jones}, {Juric},
  {Kaasalainen}, {Styliani}, {Kafka}, {Kahn}, {Kaib}, {Kalirai}, {Kantor},
  {Kasliwal}, {Keeton}, {Kessler}, {Knezevic}, {Kowalski}, {Krabbendam},
  {Krughoff}, {Kulkarni}, {Kuhlman}, {Lacy}, {Lepine}, {Liang}, {Lien}, {Lira},
  {Long}, {Lorenz}, {Lotz}, {Lupton}, {Lutz}, {Macri}, {Mahabal}, {Mandelbaum},
  {Marshall}, {May}, {McGehee}, {Meadows}, {Meert}, {Milani}, {Miller},
  {Miller}, {Mills}, {Minniti}, {Monet}, {Mukadam}, {Nakar}, {Neill}, {Newman},
  {Nikolaev}, {Nordby}, {O'Connor}, {Oguri}, {Oliver}, {Olivier}, {Olsen},
  {Olsen}, {Olszewski}, {Oluseyi}, {Padilla}, {Parker}, {Pepper}, {Peterson},
  {Petry}, {Pinto}, {Pizagno}, {Popescu}, {Prsa}, {Radcka}, {Raddick},
  {Rasmussen}, {Rau}, {Rho}, {Rhoads}, {Richards}, {Ridgway}, {Robertson},
  {Roskar}, {Saha}, {Sarajedini}, {Scannapieco}, {Schalk}, {Schindler},
  {Schmidt}, {Schmidt}, {Schneider}, {Schumacher}, {Scranton}, {Sebag},
  {Seppala}, {Shemmer}, {Simon}, {Sivertz}, {Smith}, {Allyn Smith}, {Smith},
  {Spitz}, {Stanford}, {Stassun}, {Strader}, {Strauss}, {Stubbs}, {Sweeney},
  {Szalay}, {Szkody}, {Takada}, {Thorman}, {Trilling}, {Trimble}, {Tyson}, {Van
  Berg}, {Vanden Berk}, {VanderPlas}, {Verde}, {Vrsnak}, {Walkowicz},
  {Wandelt}, {Wang}, {Wang}, {Warner}, {Wechsler}, {West}, {Wiecha},
  {Williams}, {Willman}, {Wittman}, {Wolff}, {Wood-Vasey}, {Wozniak}, {Young},
  {Zentner}, \& {Zhan}}]{LSST-science-book}
{Abell}, P.~A., {Allison}, J., {Anderson}, S.~F., {et~al.} 2009, arXiv
  e-prints, arXiv:0912.0201.
\newblock \doarXiv{0912.0201}

\bibitem[{Ade {et~al.}(2019)}]{SimonsObservatory}
Ade, P., {et~al.} 2019, JCAP, 02, 056, \dodoi{10.1088/1475-7516/2019/02/056}

\bibitem[{{Aghanim} {et~al.}(2018){Aghanim}, {Akrami}, {Ashdown}, {Aumont},
  {Baccigalupi}, {Ballardini}, {Banday}, {Barreiro}, {Bartolo}, {Basak},
  {Benabed}, {Bernard}, {Bersanelli}, {Bielewicz}, {Bock}, {Bond}, {Borrill},
  {Bouchet}, {Boulanger}, {Bucher}, {Burigana}, {Calabrese}, {Cardoso},
  {Carron}, {Challinor}, {Chiang}, {Colombo}, {Combet}, {Crill}, {Cuttaia}, {de
  Bernardis}, {de Zotti}, {Delabrouille}, {Di Valentino}, {Diego}, {Dor{\'e}},
  {Douspis}, {Ducout}, {Dupac}, {Efstathiou}, {Elsner}, {En{\ss}lin},
  {Eriksen}, {Fantaye}, {Fernandez-Cobos}, {Forastieri}, {Frailis}, {Fraisse},
  {Franceschi}, {Frolov}, {Galeotta}, {Galli}, {Ganga}, {G{\'e}nova-Santos},
  {Gerbino}, {Ghosh}, {Gonz{\'a}lez-Nuevo}, {G{\'o}rski}, {Gratton},
  {Gruppuso}, {Gudmundsson}, {Hamann}, {Hand ley}, {Hansen}, {Herranz},
  {Hivon}, {Huang}, {Jaffe}, {Jones}, {Karakci}, {Keih{\"a}nen}, {Keskitalo},
  {Kiiveri}, {Kim}, {Knox}, {Krachmalnicoff}, {Kunz}, {Kurki-Suonio},
  {Lagache}, {Lamarre}, {Lasenby}, {Lattanzi}, {Lawrence}, {Le Jeune},
  {Levrier}, {Lewis}, {Liguori}, {Lilje}, {Lindholm}, {L{\'o}pez-Caniego},
  {Lubin}, {Ma}, {Mac{\'\i}as-P{\'e}rez}, {Maggio}, {Maino}, {Mandolesi},
  {Mangilli}, {Marcos-Caballero}, {Maris}, {Martin},
  {Mart{\'\i}nez-Gonz{\'a}lez}, {Matarrese}, {Mauri}, {McEwen}, {Melchiorri},
  {Mennella}, {Migliaccio}, {Miville-Desch{\^e}nes}, {Molinari}, {Moneti},
  {Montier}, {Morgante}, {Moss}, {Natoli}, {Pagano}, {Paoletti}, {Partridge},
  {Patanchon}, {Perrotta}, {Pettorino}, {Piacentini}, {Polastri}, {Polenta},
  {Puget}, {Rachen}, {Reinecke}, {Remazeilles}, {Renzi}, {Rocha}, {Rosset},
  {Roudier}, {Rubi{\~n}o-Mart{\'\i}n}, {Ruiz-Granados}, {Salvati}, {Sandri},
  {Savelainen}, {Scott}, {Sirignano}, {Sunyaev}, {Suur-Uski}, {Tauber},
  {Tavagnacco}, {Tenti}, {Toffolatti}, {Tomasi}, {Trombetti}, {Valiviita}, {Van
  Tent}, {Vielva}, {Villa}, {Vittorio}, {Wandelt}, {Wehus}, {White}, {White},
  {Zacchei}, \& {Zonca}}]{P18:phi}
{Aghanim}, N., {Akrami}, Y., {Ashdown}, M., {et~al.} 2018, \aap

\bibitem[{Aiola {et~al.}(2020)}]{ACT:2020gnv}
Aiola, S., {et~al.} 2020, JCAP, 12, 047, \dodoi{10.1088/1475-7516/2020/12/047}

\bibitem[{{Alonso} {et~al.}(2016){Alonso}, {Louis}, {Bull}, \&
  {Ferreira}}]{Alonso2016}
{Alonso}, D., {Louis}, T., {Bull}, P., \& {Ferreira}, P.~G. 2016, \prd, 94,
  043522, \dodoi{10.1103/PhysRevD.94.043522}

\bibitem[{Amodeo {et~al.}(2021)}]{Amodeo:2020mmu}
Amodeo, S., {et~al.} 2021, Phys. Rev. D, 103, 063514,
  \dodoi{10.1103/PhysRevD.103.063514}

\bibitem[{Battaglia(2016)}]{Battaglia:2016xbi}
Battaglia, N. 2016, JCAP, 08, 058, \dodoi{10.1088/1475-7516/2016/08/058}

\bibitem[{{Battaglia} {et~al.}(2012){Battaglia}, {Bond}, {Pfrommer}, \&
  {Sievers}}]{Battaglia2012}
{Battaglia}, N., {Bond}, J.~R., {Pfrommer}, C., \& {Sievers}, J.~L. 2012, \apj,
  758, 75, \dodoi{10.1088/0004-637X/758/2/75}

\bibitem[{{Battaglia} {et~al.}(2010){Battaglia}, {Bond}, {Pfrommer}, {Sievers},
  \& {Sijacki}}]{BBPSS}
{Battaglia}, N., {Bond}, J.~R., {Pfrommer}, C., {Sievers}, J.~L., \& {Sijacki},
  D. 2010, \apj, 725, 91, \dodoi{10.1088/0004-637X/725/1/91}

\bibitem[{Battaglia {et~al.}(2015)Battaglia, Hill, \&
  Murray}]{Battaglia:2014era}
Battaglia, N., Hill, J.~C., \& Murray, N. 2015, Astrophys. J., 812, 154,
  \dodoi{10.1088/0004-637X/812/2/154}

\bibitem[{{Calafut} {et~al.}(2021){Calafut}, {Gallardo}, {Vavagiakis},
  {Amodeo}, {Aiola}, {Austermann}, {Battaglia}, {Battistelli}, {Beall}, {Bean},
  {Bond}, {Calabrese}, {Choi}, {Cothard}, {Devlin}, {Duell}, {Duff},
  {Duivenvoorden}, {Dunkley}, {Dunner}, {Ferraro}, {Guan}, {Hill}, {Hilton},
  {Hilton}, {Hlo{\v{z}}ek}, {Huber}, {Hubmayr}, {Huffenberger}, {Hughes},
  {Koopman}, {Kosowsky}, {Li}, {Lokken}, {Madhavacheril}, {McMahon}, {Moodley},
  {Naess}, {Nati}, {Newburgh}, {Niemack}, {Page}, {Partridge}, {Schaan},
  {Schillaci}, {Sif{\'o}n}, {Spergel}, {Staggs}, {Ullom}, {Vale}, {Van
  Engelen}, {Van Lanen}, {Wollack}, \& {Xu}}]{Calafut2021}
{Calafut}, V., {Gallardo}, P.~A., {Vavagiakis}, E.~M., {et~al.} 2021, \prd,
  104, 043502, \dodoi{10.1103/PhysRevD.104.043502}

\bibitem[{{Chluba} {et~al.}(2012){Chluba}, {Nagai}, {Sazonov}, \&
  {Nelson}}]{2012MNRAS.426..510C}
{Chluba}, J., {Nagai}, D., {Sazonov}, S., \& {Nelson}, K. 2012, \mnras, 426,
  510, \dodoi{10.1111/j.1365-2966.2012.21741.x}

\bibitem[{Cooray \& Sheth(2002)}]{Cooray:2002dia}
Cooray, A., \& Sheth, R.~K. 2002, Phys. Rept., 372, 1,
  \dodoi{10.1016/S0370-1573(02)00276-4}

\bibitem[{{De Bernardis} {et~al.}(2017){De Bernardis}, {Aiola}, {Vavagiakis},
  {Battaglia}, {Niemack}, {Beall}, {Becker}, {Bond}, {Calabrese}, {Cho},
  {Coughlin}, {Datta}, {Devlin}, {Dunkley}, {Dunner}, {Ferraro}, {Fox},
  {Gallardo}, {Halpern}, {Hand}, {Hasselfield}, {Henderson}, {Hill}, {Hilton},
  {Hilton}, {Hincks}, {Hlozek}, {Hubmayr}, {Huffenberger}, {Hughes}, {Irwin},
  {Koopman}, {Kosowsky}, {Li}, {Louis}, {Lungu}, {Madhavacheril}, {Maurin},
  {McMahon}, {Moodley}, {Naess}, {Nati}, {Newburgh}, {Nibarger}, {Page},
  {Partridge}, {Schaan}, {Schmitt}, {Sehgal}, {Sievers}, {Simon}, {Spergel},
  {Staggs}, {Stevens}, {Thornton}, {van Engelen}, {Van Lanen}, \&
  {Wollack}}]{DeB2017}
{De Bernardis}, F., {Aiola}, S., {Vavagiakis}, E.~M., {et~al.} 2017, \jcap,
  2017, 008, \dodoi{10.1088/1475-7516/2017/03/008}

\bibitem[{Dvorkin {et~al.}(2009)Dvorkin, Hu, \& Smith}]{Dvorkin:2009ah}
Dvorkin, C., Hu, W., \& Smith, K.~M. 2009, \prd, 79, 107302

\bibitem[{Dvorkin \& Smith(2009)}]{Dvorkin:2008:tau-est}
Dvorkin, C., \& Smith, K.~M. 2009, \prd, 79, 043003

\bibitem[{{En{\ss}lin} {et~al.}(2007){En{\ss}lin}, {Pfrommer}, {Springel}, \&
  {Jubelgas}}]{2007A&A...473...41E}
{En{\ss}lin}, T.~A., {Pfrommer}, C., {Springel}, V., \& {Jubelgas}, M. 2007,
  \aap, 473, 41, \dodoi{10.1051/0004-6361:20065294}

\bibitem[{Eriksen {et~al.}(2004)Eriksen, Banday, Gorski, \&
  Lilje}]{Eriksen:2004jg}
Eriksen, H.~K., Banday, A.~J., Gorski, K.~M., \& Lilje, P.~B. 2004, Astrophys.
  J., 612, 633, \dodoi{10.1086/422807}

\bibitem[{Feng \& Holder(2018)}]{Feng2018}
Feng, C., \& Holder, G. 2018, Phys. Rev. D, 97, 123523,
  \dodoi{10.1103/PhysRevD.97.123523}

\bibitem[{Feng \& Holder(2019)}]{Feng:2018eal}
---. 2019, Phys. Rev. D, 99, 123502, \dodoi{10.1103/PhysRevD.99.123502}

\bibitem[{{Ferraro} {et~al.}(2016){Ferraro}, {Hill}, {Battaglia}, {Liu}, \&
  {Spergel}}]{Simo2016}
{Ferraro}, S., {Hill}, J.~C., {Battaglia}, N., {Liu}, J., \& {Spergel}, D.~N.
  2016, \prd, 94, 123526, \dodoi{10.1103/PhysRevD.94.123526}

\bibitem[{Gluscevic {et~al.}(2013)Gluscevic, Kamionkowski, \&
  Hanson}]{Gluscevic:2012:tau}
Gluscevic, V., Kamionkowski, M., \& Hanson, D. 2013, \prd, 87, 047303

\bibitem[{{Greco} {et~al.}(2015){Greco}, {Hill}, {Spergel}, \&
  {Battaglia}}]{2015ApJ...808..151G}
{Greco}, J.~P., {Hill}, J.~C., {Spergel}, D.~N., \& {Battaglia}, N. 2015, \apj,
  808, 151, \dodoi{10.1088/0004-637X/808/2/151}

\bibitem[{Guzman \& Meyers(2021)}]{Guzman:2021nfk}
Guzman, E., \& Meyers, J. 2021, Phys. Rev. D, 104, 043529,
  \dodoi{10.1103/PhysRevD.104.043529}

\bibitem[{Hanany {et~al.}(2019)}]{Hanany:2019lle}
Hanany, S., {et~al.} 2019.
\newblock \doarXiv{1902.10541}

\bibitem[{Hill {et~al.}(2015)Hill, Battaglia, Chluba, Ferraro, Schaan, \&
  Spergel}]{Hill:2015}
Hill, J.~C., Battaglia, N., Chluba, J., {et~al.} 2015, Phys. Rev. Lett., 115,
  261301, \dodoi{10.1103/PhysRevLett.115.261301}

\bibitem[{{Hill} {et~al.}(2016){Hill}, {Ferraro}, {Battaglia}, {Liu}, \&
  {Spergel}}]{Hill2016}
{Hill}, J.~C., {Ferraro}, S., {Battaglia}, N., {Liu}, J., \& {Spergel}, D.~N.
  2016, \prl, 117, 051301, \dodoi{10.1103/PhysRevLett.117.051301}

\bibitem[{Hill \& Pajer(2013)}]{Hill:2013}
Hill, J.~C., \& Pajer, E. 2013, Phys. Rev. D, 88, 063526,
  \dodoi{10.1103/PhysRevD.88.063526}

\bibitem[{Hill \& Spergel(2014)}]{Hill:2013dxa}
Hill, J.~C., \& Spergel, D.~N. 2014, \jcap, 02, 030

\bibitem[{Hojjati {et~al.}(2017)}]{Hojjati:2016nbx}
Hojjati, A., {et~al.} 2017, Mon. Not. Roy. Astron. Soc., 471, 1565,
  \dodoi{10.1093/mnras/stx1659}

\bibitem[{{Hurier} {et~al.}(2015){Hurier}, {Douspis}, {Aghanim},
  {Pointecouteau}, {Diego}, \& {Macias-Perez}}]{2015A&A...576A..90H}
{Hurier}, G., {Douspis}, M., {Aghanim}, N., {et~al.} 2015, \aap, 576, A90,
  \dodoi{10.1051/0004-6361/201425555}

\bibitem[{{Jubelgas} {et~al.}(2008){Jubelgas}, {Springel}, {En{\ss}lin}, \&
  {Pfrommer}}]{2008A&A...481...33J}
{Jubelgas}, M., {Springel}, V., {En{\ss}lin}, T., \& {Pfrommer}, C. 2008, \aap,
  481, 33, \dodoi{10.1051/0004-6361:20065295}

\bibitem[{Knox(1995)}]{Knox:1995dq}
Knox, L. 1995, Phys. Rev. D, 52, 4307, \dodoi{10.1103/PhysRevD.52.4307}

\bibitem[{Komatsu \& Kitayama(1999)}]{Komatsu1999}
Komatsu, E., \& Kitayama, T. 1999, Astrophys. J. Lett., 526, L1,
  \dodoi{10.1086/312364}

\bibitem[{Koukoufilippas {et~al.}(2020)Koukoufilippas, Alonso, Bilicki, \&
  Peacock}]{Koukoufilippas:2019ilu}
Koukoufilippas, N., Alonso, D., Bilicki, M., \& Peacock, J.~A. 2020, Mon. Not.
  Roy. Astron. Soc., 491, 5464, \dodoi{10.1093/mnras/stz3351}

\bibitem[{{Kusiak} {et~al.}(2021){Kusiak}, {Bolliet}, {Ferraro}, {Hill}, \&
  {Krolewski}}]{Kusiak2021}
{Kusiak}, A., {Bolliet}, B., {Ferraro}, S., {Hill}, J.~C., \& {Krolewski}, A.
  2021, \prd, 104, 043518, \dodoi{10.1103/PhysRevD.104.043518}

\bibitem[{Lewis {et~al.}(2000)Lewis, Challinor, \& Lasenby}]{Lewis:1999bs}
Lewis, A., Challinor, A., \& Lasenby, A. 2000, Astrophys. J., 538, 473,
  \dodoi{10.1086/309179}

\bibitem[{Madhavacheril {et~al.}(2019)Madhavacheril, Battaglia, Smith, \&
  Sievers}]{Madhavacheril:2019buy}
Madhavacheril, M.~S., Battaglia, N., Smith, K.~M., \& Sievers, J.~L. 2019,
  Phys. Rev. D, 100, 103532, \dodoi{10.1103/PhysRevD.100.103532}

\bibitem[{Meerburg {et~al.}(2013)Meerburg, Dvorkin, \& Spergel}]{Meerburg2013}
Meerburg, P.~D., Dvorkin, C., \& Spergel, D.~N. 2013, Astrophys. J., 779, 124,
  \dodoi{10.1088/0004-637X/779/2/124}

\bibitem[{{Meinke} {et~al.}(2021){Meinke}, {B{\"o}ckmann}, {Cohen}, {Mauskopf},
  {Scannapieco}, {Sarmento}, {Lunde}, \& {Cottle}}]{2021ApJ...913...88M}
{Meinke}, J., {B{\"o}ckmann}, K., {Cohen}, S., {et~al.} 2021, \apj, 913, 88,
  \dodoi{10.3847/1538-4357/abf2b4}

\bibitem[{{Moser} {et~al.}(2021){Moser}, {Amodeo}, {Battaglia}, {Alvarez},
  {Ferraro}, \& {Schaan}}]{Moser2021}
{Moser}, E., {Amodeo}, S., {Battaglia}, N., {et~al.} 2021, \apj, 919, 2,
  \dodoi{10.3847/1538-4357/ac0cea}

\bibitem[{Namikawa(2018)}]{Namikawa:2017:plktau}
Namikawa, T. 2018, \prd, 97, 063505

\bibitem[{Namikawa {et~al.}(2013)Namikawa, Hanson, \&
  Takahashi}]{Namikawa:2012:bhe}
Namikawa, T., Hanson, D., \& Takahashi, R. 2013, \mnras, 431, 609,
  \dodoi{10.1093/mnras/stt195}

\bibitem[{{Namikawa} {et~al.}(2021){Namikawa}, {Roy}, {Sherwin}, {Battaglia},
  \& {Spergel}}]{Namikawa:2021zhh}
{Namikawa}, T., {Roy}, A., {Sherwin}, B.~D., {Battaglia}, N., \& {Spergel},
  D.~N. 2021, arXiv e-prints, arXiv:2102.00975.
\newblock \doarXiv{2102.00975}

\bibitem[{Navarro {et~al.}(1996)Navarro, Frenk, \& White}]{NFW}
Navarro, J.~F., Frenk, C.~S., \& White, S. D.~M. 1996, Astrophys. J., 462, 563,
  \dodoi{10.1086/177173}

\bibitem[{{Nielsen} {et~al.}(2015){Nielsen}, {Churchill}, {Kacprzak}, {Murphy},
  \& {Evans}}]{2015ApJ...812...83N}
{Nielsen}, N.~M., {Churchill}, C.~W., {Kacprzak}, G.~G., {Murphy}, M.~T., \&
  {Evans}, J.~L. 2015, \apj, 812, 83, \dodoi{10.1088/0004-637X/812/1/83}

\bibitem[{{Nozawa} {et~al.}(2006){Nozawa}, {Itoh}, {Suda}, \&
  {Ohhata}}]{Sz_relativistic}
{Nozawa}, S., {Itoh}, N., {Suda}, Y., \& {Ohhata}, Y. 2006, Nuovo Cimento B
  Serie, 121, 487, \dodoi{10.1393/ncb/i2005-10223-0}

\bibitem[{Pandey {et~al.}(2020)Pandey, Baxter, \& Hill}]{Pandey:2019}
Pandey, S., Baxter, E.~J., \& Hill, J.~C. 2020, Phys. Rev. D, 101, 043525,
  \dodoi{10.1103/PhysRevD.101.043525}

\bibitem[{Pandey {et~al.}(2021)}]{Pandey:2021bdj}
Pandey, S., {et~al.} 2021.
\newblock \doarXiv{2108.01601}

\bibitem[{{Pfrommer} {et~al.}(2006){Pfrommer}, {Springel}, {En{\ss}lin}, \&
  {Jubelgas}}]{2006MNRAS.367..113P}
{Pfrommer}, C., {Springel}, V., {En{\ss}lin}, T.~A., \& {Jubelgas}, M. 2006,
  \mnras, 367, 113, \dodoi{10.1111/j.1365-2966.2005.09953.x}

\bibitem[{{Planck Collaboration} {et~al.}(2016){Planck Collaboration}, {Ade},
  {Aghanim}, {Alves}, {Arnaud}, {Ashdown}, {Aumont}, {Baccigalupi}, {Banday},
  {Barreiro}, {Bartlett}, {Bartolo}, {Battaner}, {Benabed}, {Beno{\^\i}t},
  {Benoit-L{\'e}vy}, {Bernard}, {Bersanelli}, {Bielewicz}, {Bock}, {Bonaldi},
  {Bonavera}, {Bond}, {Borrill}, {Bouchet}, {Boulanger}, {Bucher}, {Burigana},
  {Butler}, {Calabrese}, {Cardoso}, {Catalano}, {Challinor}, {Chamballu},
  {Chary}, {Chiang}, {Christensen}, {Colombi}, {Colombo}, {Combet}, {Couchot},
  {Coulais}, {Crill}, {Curto}, {Cuttaia}, {Danese}, {Davies}, {Davis}, {de
  Bernardis}, {de Rosa}, {de Zotti}, {Delabrouille}, {Delouis}, {D{\'e}sert},
  {Dickinson}, {Diego}, {Dole}, {Donzelli}, {Dor{\'e}}, {Douspis}, {Ducout},
  {Dupac}, {Efstathiou}, {Elsner}, {En{\ss}lin}, {Eriksen}, {Falgarone},
  {Fergusson}, {Finelli}, {Forni}, {Frailis}, {Fraisse}, {Franceschi},
  {Frejsel}, {Galeotta}, {Galli}, {Ganga}, {Ghosh}, {Giard},
  {Giraud-H{\'e}raud}, {Gjerl{\o}w}, {Gonz{\'a}lez-Nuevo}, {G{\'o}rski},
  {Gratton}, {Gregorio}, {Gruppuso}, {Gudmundsson}, {Hansen}, {Hanson},
  {Harrison}, {Helou}, {Henrot-Versill{\'e}}, {Hern{\'a}ndez-Monteagudo},
  {Herranz}, {Hildebrandt}, {Hivon}, {Hobson}, {Holmes}, {Hornstrup}, {Hovest},
  {Huffenberger}, {Hurier}, {Jaffe}, {Jaffe}, {Jones}, {Juvela},
  {Keih{\"a}nen}, {Keskitalo}, {Kisner}, {Kneissl}, {Knoche}, {Kunz},
  {Kurki-Suonio}, {Lagache}, {L{\"a}hteenm{\"a}ki}, {Lamarre}, {Lasenby},
  {Lattanzi}, {Lawrence}, {Leahy}, {Leonardi}, {Lesgourgues}, {Levrier},
  {Liguori}, {Lilje}, {Linden-V{\o}rnle}, {L{\'o}pez-Caniego}, {Lubin},
  {Mac{\'\i}as-P{\'e}rez}, {Maggio}, {Maino}, {Mandolesi}, {Mangilli}, {Maris},
  {Marshall}, {Martin}, {Mart{\'\i}nez-Gonz{\'a}lez}, {Masi}, {Matarrese},
  {McGehee}, {Meinhold}, {Melchiorri}, {Mendes}, {Mennella}, {Migliaccio},
  {Mitra}, {Miville-Desch{\^e}nes}, {Moneti}, {Montier}, {Morgante},
  {Mortlock}, {Moss}, {Munshi}, {Murphy}, {Nati}, {Natoli}, {Netterfield},
  {N{\o}rgaard-Nielsen}, {Noviello}, {Novikov}, {Novikov}, {Orlando},
  {Oxborrow}, {Paci}, {Pagano}, {Pajot}, {Paladini}, {Paoletti}, {Partridge},
  {Pasian}, {Patanchon}, {Pearson}, {Peel}, {Perdereau}, {Perotto}, {Perrotta},
  {Pettorino}, {Piacentini}, {Piat}, {Pierpaoli}, {Pietrobon}, {Plaszczynski},
  {Pointecouteau}, {Polenta}, {Pratt}, {Pr{\'e}zeau}, {Prunet}, {Puget},
  {Rachen}, {Reach}, {Rebolo}, {Reinecke}, {Remazeilles}, {Renault}, {Renzi},
  {Ristorcelli}, {Rocha}, {Rosset}, {Rossetti}, {Roudier},
  {Rubi{\~n}o-Mart{\'\i}n}, {Rusholme}, {Sandri}, {Santos}, {Savelainen},
  {Savini}, {Scott}, {Seiffert}, {Shellard}, {Spencer}, {Stolyarov}, {Stompor},
  {Strong}, {Sudiwala}, {Sunyaev}, {Sutton}, {Suur-Uski}, {Sygnet}, {Tauber},
  {Terenzi}, {Toffolatti}, {Tomasi}, {Tristram}, {Tucci}, {Tuovinen}, {Umana},
  {Valenziano}, {Valiviita}, {Van Tent}, {Vidal}, {Vielva}, {Villa}, {Wade},
  {Wandelt}, {Watson}, {Wehus}, {Wilkinson}, {Yvon}, {Zacchei}, \&
  {Zonca}}]{2016A&A...594A..25P}
{Planck Collaboration}, {Ade}, P.~A.~R., {Aghanim}, N., {et~al.} 2016, \aap,
  594, A25, \dodoi{10.1051/0004-6361/201526803}

\bibitem[{{Planck Collaboration} {et~al.}(2020){Planck Collaboration},
  {Aghanim}, {Akrami}, {Ashdown}, {Aumont}, {Baccigalupi}, {Ballardini},
  {Banday}, {Barreiro}, {Bartolo}, {Basak}, {Battye}, {Benabed}, {Bernard},
  {Bersanelli}, {Bielewicz}, {Bock}, {Bond}, {Borrill}, {Bouchet}, {Boulanger},
  {Bucher}, {Burigana}, {Butler}, {Calabrese}, {Cardoso}, {Carron},
  {Challinor}, {Chiang}, {Chluba}, {Colombo}, {Combet}, {Contreras}, {Crill},
  {Cuttaia}, {de Bernardis}, {de Zotti}, {Delabrouille}, {Delouis}, {Di
  Valentino}, {Diego}, {Dor{\'e}}, {Douspis}, {Ducout}, {Dupac}, {Dusini},
  {Efstathiou}, {Elsner}, {En{\ss}lin}, {Eriksen}, {Fantaye}, {Farhang},
  {Fergusson}, {Fernandez-Cobos}, {Finelli}, {Forastieri}, {Frailis},
  {Fraisse}, {Franceschi}, {Frolov}, {Galeotta}, {Galli}, {Ganga},
  {G{\'e}nova-Santos}, {Gerbino}, {Ghosh}, {Gonz{\'a}lez-Nuevo}, {G{\'o}rski},
  {Gratton}, {Gruppuso}, {Gudmundsson}, {Hamann}, {Handley}, {Hansen},
  {Herranz}, {Hildebrandt}, {Hivon}, {Huang}, {Jaffe}, {Jones}, {Karakci},
  {Keih{\"a}nen}, {Keskitalo}, {Kiiveri}, {Kim}, {Kisner}, {Knox},
  {Krachmalnicoff}, {Kunz}, {Kurki-Suonio}, {Lagache}, {Lamarre}, {Lasenby},
  {Lattanzi}, {Lawrence}, {Le Jeune}, {Lemos}, {Lesgourgues}, {Levrier},
  {Lewis}, {Liguori}, {Lilje}, {Lilley}, {Lindholm}, {L{\'o}pez-Caniego},
  {Lubin}, {Ma}, {Mac{\'\i}as-P{\'e}rez}, {Maggio}, {Maino}, {Mandolesi},
  {Mangilli}, {Marcos-Caballero}, {Maris}, {Martin}, {Martinelli},
  {Mart{\'\i}nez-Gonz{\'a}lez}, {Matarrese}, {Mauri}, {McEwen}, {Meinhold},
  {Melchiorri}, {Mennella}, {Migliaccio}, {Millea}, {Mitra},
  {Miville-Desch{\^e}nes}, {Molinari}, {Montier}, {Morgante}, {Moss}, {Natoli},
  {N{\o}rgaard-Nielsen}, {Pagano}, {Paoletti}, {Partridge}, {Patanchon},
  {Peiris}, {Perrotta}, {Pettorino}, {Piacentini}, {Polastri}, {Polenta},
  {Puget}, {Rachen}, {Reinecke}, {Remazeilles}, {Renzi}, {Rocha}, {Rosset},
  {Roudier}, {Rubi{\~n}o-Mart{\'\i}n}, {Ruiz-Granados}, {Salvati}, {Sandri},
  {Savelainen}, {Scott}, {Shellard}, {Sirignano}, {Sirri}, {Spencer},
  {Sunyaev}, {Suur-Uski}, {Tauber}, {Tavagnacco}, {Tenti}, {Toffolatti},
  {Tomasi}, {Trombetti}, {Valenziano}, {Valiviita}, {Van Tent}, {Vibert},
  {Vielva}, {Villa}, {Vittorio}, {Wandelt}, {Wehus}, {White}, {White},
  {Zacchei}, \& {Zonca}}]{P18:main}
{Planck Collaboration}, {Aghanim}, N., {Akrami}, Y., {et~al.} 2020, \aap, 641,
  A6, \dodoi{10.1051/0004-6361/201833910}

\bibitem[{Reichardt {et~al.}(2021)}]{Reichardt:2020}
Reichardt, C.~L., {et~al.} 2021, Astrophys. J., 908, 199,
  \dodoi{10.3847/1538-4357/abd407}

\bibitem[{Remazeilles {et~al.}(2011)Remazeilles, Delabrouille, \&
  Cardoso}]{Remazeilles:2010hq}
Remazeilles, M., Delabrouille, J., \& Cardoso, J.-F. 2011, Mon. Not. Roy.
  Astron. Soc., 410, 2481, \dodoi{10.1111/j.1365-2966.2010.17624.x}

\bibitem[{Roy {et~al.}(2021)Roy, Kulkarni, Meerburg, Challinor, Baccigalupi,
  Lapi, \& Haehnelt}]{Roy:2020cqn}
Roy, A., Kulkarni, G., Meerburg, P.~D., {et~al.} 2021, JCAP, 01, 003,
  \dodoi{10.1088/1475-7516/2021/01/003}

\bibitem[{Roy {et~al.}(2018)Roy, Lapi, Spergel, \& Baccigalupi}]{Roy:2018}
Roy, A., Lapi, A., Spergel, D., \& Baccigalupi, C. 2018, \jcap, 05, 014,
  \dodoi{10.1088/1475-7516/2018/05/014}

\bibitem[{Roy {et~al.}(2020)Roy, Lapi, Spergel, Basak, \&
  Baccigalupi}]{Roy:2019qsl}
Roy, A., Lapi, A., Spergel, D., Basak, S., \& Baccigalupi, C. 2020, JCAP, 03,
  062, \dodoi{10.1088/1475-7516/2020/03/062}

\bibitem[{{Schaan} {et~al.}(2016){Schaan}, {Ferraro}, {Vargas-Maga{\~n}a},
  {Smith}, {Ho}, {Aiola}, {Battaglia}, {Bond}, {De Bernardis}, {Calabrese},
  {Cho}, {Devlin}, {Dunkley}, {Gallardo}, {Hasselfield}, {Henderson}, {Hill},
  {Hincks}, {Hlozek}, {Hubmayr}, {Hughes}, {Irwin}, {Koopman}, {Kosowsky},
  {Li}, {Louis}, {Lungu}, {Madhavacheril}, {Maurin}, {McMahon}, {Moodley},
  {Naess}, {Nati}, {Newburgh}, {Niemack}, {Page}, {Pappas}, {Partridge},
  {Schmitt}, {Sehgal}, {Sherwin}, {Sievers}, {Spergel}, {Staggs}, {van
  Engelen}, {Wollack}, \& {ACTPol Collaboration}}]{2016PhRvD..93h2002S}
{Schaan}, E., {Ferraro}, S., {Vargas-Maga{\~n}a}, M., {et~al.} 2016, \prd, 93,
  082002, \dodoi{10.1103/PhysRevD.93.082002}

\bibitem[{Schaan {et~al.}(2021)}]{AtacamaCosmologyTelescope:2020wtv}
Schaan, E., {et~al.} 2021, Phys. Rev. D, 103, 063513,
  \dodoi{10.1103/PhysRevD.103.063513}

\bibitem[{Scoccimarro {et~al.}(2001)Scoccimarro, Sheth, Hui, \&
  Jain}]{Scoccimarro2000}
Scoccimarro, R., Sheth, R.~K., Hui, L., \& Jain, B. 2001, Astrophys. J., 546,
  20, \dodoi{10.1086/318261}

\bibitem[{{Shao} \& {Fang}(2016)}]{Shao2016}
{Shao}, J., \& {Fang}, T. 2016, \mnras, 458, 3773, \dodoi{10.1093/mnras/stw501}

\bibitem[{Singh {et~al.}(2016)Singh, Majumdar, Nath, Refregier, \&
  Silk}]{Singh:2015}
Singh, P., Majumdar, S., Nath, B.~B., Refregier, A., \& Silk, J. 2016, Mon.
  Not. Roy. Astron. Soc., 456, 1495, \dodoi{10.1093/mnras/stv2750}

\bibitem[{Smith {et~al.}(2012)Smith, Hanson, LoVerde, Hirata, \&
  Zahn}]{Smith:2010gu}
Smith, K.~M., Hanson, D., LoVerde, M., Hirata, C.~M., \& Zahn, O. 2012, \jcap,
  06, 014

\bibitem[{Sobrin {et~al.}(2021)}]{SPT-3G:2021vps}
Sobrin, J.~A., {et~al.} 2021.
\newblock \doarXiv{2106.11202}

\bibitem[{{Springel}(2005)}]{Gadget}
{Springel}, V. 2005, \mnras, 364, 1105,
  \dodoi{10.1111/j.1365-2966.2005.09655.x}

\bibitem[{{Springel} \& {Hernquist}(2003)}]{SpHr2003}
{Springel}, V., \& {Hernquist}, L. 2003, \mnras, 339, 289,
  \dodoi{10.1046/j.1365-8711.2003.06206.x}

\bibitem[{{Su} {et~al.}(2011){Su}, {Yadav}, {McQuinn}, {Yoo}, \&
  {Zaldarriaga}}]{Su:2011ff}
{Su}, M., {Yadav}, A. P.~S., {McQuinn}, M., {Yoo}, J., \& {Zaldarriaga}, M.
  2011, arXiv e-prints

\bibitem[{{Sunyaev} \& {Zeldovich}(1970)}]{Sunyaev1970}
{Sunyaev}, R.~A., \& {Zeldovich}, Y.~B. 1970, \apss, 7, 3,
  \dodoi{10.1007/BF00653471}

\bibitem[{{Sunyaev} \& {Zeldovich}(1972)}]{Sunyaev1972}
---. 1972, Comments on Astrophysics and Space Physics, 4, 173

\bibitem[{Tinker {et~al.}(2008)Tinker, Kravtsov, Klypin, Abazajian, Warren,
  Yepes, Gottlober, \& Holz}]{Tinker:2008}
Tinker, J.~L., Kravtsov, A.~V., Klypin, A., {et~al.} 2008, Astrophys. J., 688,
  709, \dodoi{10.1086/591439}

\bibitem[{{Tumlinson} {et~al.}(2017){Tumlinson}, {Peeples}, \&
  {Werk}}]{2017ARA&A..55..389T}
{Tumlinson}, J., {Peeples}, M.~S., \& {Werk}, J.~K. 2017, \araa, 55, 389,
  \dodoi{10.1146/annurev-astro-091916-055240}

\bibitem[{{Vavagiakis} {et~al.}(2021){Vavagiakis}, {Gallardo}, {Calafut},
  {Amodeo}, {Aiola}, {Austermann}, {Battaglia}, {Battistelli}, {Beall}, {Bean},
  {Bond}, {Calabrese}, {Choi}, {Cothard}, {Devlin}, {Duell}, {Duff},
  {Duivenvoorden}, {Dunkley}, {Dunner}, {Ferraro}, {Guan}, {Hill}, {Hilton},
  {Hilton}, {Hlo{\v{z}}ek}, {Huber}, {Hubmayr}, {Huffenberger}, {Hughes},
  {Koopman}, {Kosowsky}, {Li}, {Lokken}, {Madhavacheril}, {McMahon}, {Moodley},
  {Naess}, {Nati}, {Newburgh}, {Niemack}, {Page}, {Partridge}, {Schaan},
  {Schillaci}, {Sif{\'o}n}, {Spergel}, {Staggs}, {Ullom}, {Vale}, {Van
  Engelen}, {Van Lanen}, {Wollack}, \& {Xu}}]{Vava2021}
{Vavagiakis}, E.~M., {Gallardo}, P.~A., {Calafut}, V., {et~al.} 2021, \prd,
  104, 043503, \dodoi{10.1103/PhysRevD.104.043503}

\bibitem[{Werk {et~al.}(2014)}]{Werk:2014fza}
Werk, J.~K., {et~al.} 2014, Astrophys. J., 792, 8,
  \dodoi{10.1088/0004-637X/792/1/8}

\bibitem[{Yan {et~al.}(2021)}]{Yan:2021gfo}
Yan, Z., {et~al.} 2021, Astron. Astrophys., 651, A76,
  \dodoi{10.1051/0004-6361/202140568}

\bibitem[{{Zhao}(1996)}]{Zhao1996}
{Zhao}, H. 1996, \mnras, 278, 488, \dodoi{10.1093/mnras/278.2.488}

\bibitem[{Zheng {et~al.}(2005)Zheng, Berlind, Weinberg, Benson, Baugh, Cole,
  Dave, Frenk, Katz, \& Lacey}]{Zheng2004}
Zheng, Z., Berlind, A.~A., Weinberg, D.~H., {et~al.} 2005, Astrophys. J., 633,
  791, \dodoi{10.1086/466510}

\end{thebibliography}
\bibliographystyle{aasjournal}

\end{document}